\documentclass[aps,prx]{revtex4}

\usepackage{amsfonts}
\usepackage{amsmath}
\usepackage{amsbsy}
\usepackage{amssymb}
\usepackage{amsthm}
\usepackage{array} 
\usepackage{float}
\usepackage{multirow}
\usepackage{graphicx}
\usepackage[caption=false]{subfig}

\usepackage[usenames, dvipsnames]{color} 
\usepackage{xcolor}

\usepackage{hyperref}
\usepackage{mathtools}

\usepackage{placeins} 

\begin{document}

\preprint{}

\title{Transition to anomalous dynamics in a simple random map}

\author{Jin Yan\footnote[1]{Email: yan@wias-berlin.de}}
\affiliation{Weierstrass Institute for Applied Analysis and Stochastics, Mohrenstr. 39, 10117 Berlin, Germany}
\affiliation{Max Planck Institute for the Physics of Complex Systems, N\"othnitzer Str.\ 38, 01187 Dresden, Germany}
\author{Moitrish Majumdar}
\affiliation{Department of Applied Mathematics, University of California, Merced, 5200 N.\ Lake Road, 95343 Merced, California, United States of America}
\author{Stefano Ruffo}
\affiliation{SISSA, Via Bonomea 265, 34136 Trieste, Italy}
\affiliation{INFN Sezione di Trieste, via Valerio, 2 34127 Trieste, Italy}
\affiliation{Istituto dei Sistemi Complessi, via Madonna del Piano 10 - 50019 Sesto Fiorentino, Italy}
\author{Yuzuru Sato}
\affiliation{RIES/Department of Mathematics, Hokkaido University, N12 W7 Kita-ku, Sapporo, 0600812 Hokkaido, Japan}
\affiliation{London Mathematical Laboratory, 8 Margravine Gardens, London W6 8RH, United Kingdom}
\author{Christian Beck}
\affiliation{School of Mathematical Sciences, Queen Mary University of London, Mile End Road, London E1 4NS, United Kingdom}
\affiliation{The Alan Turing Institute, 96 Euston Road, London NW1 2DB, United Kingdom}
\author{Rainer Klages}
\affiliation{School of Mathematical Sciences, Queen Mary University of London, Mile End Road, London E1 4NS, United Kingdom}
\affiliation{London Mathematical Laboratory, 8 Margravine Gardens, London W6 8RH, United Kingdom}

\date{\today}

\begin{abstract}

  The famous {doubling map} (or dyadic transformation) is perhaps the
  simplest deterministic dynamical system exhibiting chaotic
  dynamics. It is a piecewise linear time-discrete map on the unit
  interval with a uniform slope larger than one, hence expanding, with
  a positive Lyapunov exponent and a uniform invariant density. If the
  slope is less than one the map becomes contracting, the Lyapunov
  exponent is negative, and the density trivially collapses onto a
  fixed point. Sampling from these two different types of maps at each
  time step by randomly selecting the expanding one with probability
  $p$, and the contracting one with probability $1-p$, gives a
  prototype of a random dynamical system. Here we calculate the
  invariant density of this simple random map, as well as its position
  autocorrelation function, analytically and numerically under
  variation of $p$. We find that the map exhibits a non-trivial
  transition from fully chaotic to completely regular dynamics by
  generating a long-time anomalous dynamics at a critical sampling
  probability $p_c$, defined by a zero Lyapunov exponent. This
  anomalous dynamics is characterised by an infinte invariant density,
  weak ergodicity breaking and power law correlation decay.
  
\end{abstract}

\pacs{05.45.Ac, 05.45.Df, 05.60.Cd (check!)}

\maketitle

\begin{quotation}

{\bf A random dynamical system consists of a setting in which
  different types of dynamics are generated randomly in time. Here we
  consider a simple example, where an irregular or a regular dynamics
  is randomly selected at each discrete time step for determining the
  outcome at the next time step. By varying the sampling probability
  to select the one or the other dynamics, the completely irregular,
  respectively regular dynamics are recovered as limiting
  cases. Hence, there must be a transition between these two
  states. In our paper we explore this transition in detail. We find
  that there is a critical sampling probability at which the dynamics
  becomes anomalous by exhibiting a certain type of intermittency,
  i.e., a specific mixing between regular and irregular dynamics.}
  
\end{quotation}
  

\section{Introduction}

Dynamical systems theory and the theory of stochastic processes
generically refer to rather different fields of research. The former
focuses on understanding the nonlinear and chaotic properties of
typically deterministic dynamical systems, where the dynamics is
completely determined by the initial conditions and the given
equations of motion \cite{KaHa95,Ott}. Stochastic processes, in
contrast, are governed by equations of motion containing random
variables, which yields different outcomes when starting from the same
initial conditions \cite{vK,Gard09}. Interestingly, deterministic
dynamical systems may exhibit chaotic properties that, on a certain
scale, can well be reproduced by a matching to stochastic processes
\cite{LaMa,Beck}. This is due to dynamical instabilities in the
deterministic equations of motion leading to exponential separation of
trajectories starting from nearby initial conditions
\cite{KaHa95,Ott}. The resulting cross-link between deterministic and
stochastic dynamics has been exploited particularly for understanding
the emergence of irreversibility, transport and complexity on
macroscopic scales in nonequilibrium statistical physics, starting
from deterministic chaos in microscopic equations of motion
\cite{EvMo90,HoB99,Gasp,Do99,Kla06}. This approach, which employs a
suitable coarse-graining \cite{CFLV08}, amends a broad spectrum of
conventional methods in nonequilibrium statistical physics, where
stochastic randomness is put in by hand on a more microscopic level
already \cite{Reif,Reichl16}. 

Another combination of deterministic and stochastic dynamics consists
of what is called a random dynamical system
\cite{Kifer86,Kap90,Arn98,LaMa}. According to Ref.~\cite{Kifer86}, the
origin of this field of research traces back to a short note by Ulam
and von Neumann in 1945 about random ergodic theorems \cite{UlvN45}
that triggered a number of pioneering works. In random dynamical
systems the equations of motion include random variables but are
analysed within the framework of dynamical systems theory and ergodic
theory. One might assume that all such systems can be understood by
applying random perturbation theory \cite{FrWe12} to the underlying
deterministic dynamics. However, in general the randomness may not
necessarily be small and could consist of any type of stochastic
dynamics. Accordingly, it has been found that random dynamical systems
may exhibit completely novel dynamical properties.  {Most
  well-known examples are noise-induced chaos
  \cite{mayer1981influence, crutchfield1982fluctuations} and
  noise-induced order \cite{MaTs83,Mats84}, where adding external
  noise to a certain class of one-dimensional (potentially) chaotic
  maps induces a transition from an ordered state to chaos, or from
  chaos to an ordered state, changing the sign of the top Lyapunov
  exponent and the positivity of the Kolmogorov-Sinai entropy. More
  recently, a computer-aided proof has been provided for these
  phenomena \cite{GaMoNi20,CSNG22}. In these classes of random maps
  which exhibit noise-induced transitions, intermittent behaviour may
  be observed at the zero-crossing point of the top Lyapunov exponent
  similar to the Pelikan map. Other important examples are suppression
  and enhancement of chaotic diffusion by noise
  \cite{RKla01a,RKla01b},} random attractors
\cite{RCDGK14,FSS17,chekroun2011stochastic} and their stochastic
bifurcations \cite{SDLR19,ELR19}, as well as intricate statistical and
ergodic properties
\cite{Pel84,lasota1987noise,Blank01,abbasi2018iterated,SaKl19,maldonado2021phase,KaZe23,HaYa23,Nis23}
requiring a novel mathematical language for their description
\cite{CDLR17,ELR19b}. This new theory has important applications to,
e.g., fluid dynamics \cite{LOC90,YOC91,FSS17}, climate science
\cite{chekroun2011stochastic,van2012origin}, and neural dynamics
\cite{lin2008shear,lin2009reliability}

More recently, it was shown that simple random dynamical systems can
generate anomalous diffusion that matches to subdiffusive continuous
time random walks \cite{SaKl19}, which continues a theoretical physics
line of research starting from blow-out bifurcations and on-off
intermittency
\cite{Pik84,FuYa85,FuYa86,LOC90,YOC91,PiGr91,Pik92,PST93,HPH94,OtSo94,HaMi97,HHF99,MHB01}.
Anomalous diffusion means that the mean square displacement (MSD) of
an ensemble of particles increases nonlinearly in time, $\langle
x^2\rangle\sim t^{\alpha}$ with $\alpha\neq1$, while for normal
diffusion $\alpha=1$. This connects random dynamical systems theory
with yet another big field of research, which is the theory of
anomalous stochastic processes and their corresponding anomalous
transport properties \cite{MeKl00,KRS08,HoFr13,MJCB14,ZDK15}. We
remark that anomalous dynamics and anomalous transport are known to
occur in deterministic dynamical systems
\cite{KRS08,ZGNR86,SZK93,KSZ96,Zas02,GeTo84,ZuKl93a,ArCr03,Bar03,KCKSG06,KKCSG06,Kla06}
if they are weakly chaotic \cite{ZaUs01,Gala03,Kla06,Kla13}, that is,
there is still sensitivity to initial conditions, but the dynamical
instability is weaker than exponential. In addition to this fully
deterministic origin of anomalous dynamics, random dynamical systems
thus provide a new access road to understand the emergence of
anomalous dynamical properties at the interface between the theory of
dynamical systems and stochastic theory \cite{MHB01,RCDGK14,SaKl19}.

A strikingly simple random dynamical system exhibiting anomalous
dynamical properties was put forward in the paper by Pelikan
Ref.~\cite{Pel84}. Therein sufficient mathematical conditions were
derived under which the invariant density of random maps is absolutely
continuous. These theorems were illustrated by a specific example, a
piecewise linear time-discrete random map on the unit interval that we
call the Pelikan map, also studied in Refs.~\cite{LOC90,AAN98,HaYa23};
see Eqs.~\eqref{rdsT} given later and the respective discussion. This
map displays a transition of the corresponding invariant density,
under variation of a control parameter, from a uniform density,
generated by chaotic dynamics with a positive Lyapunov exponent, to a
density contracting onto a fixed point, associated with a negative
Lyapunov exponent. For this map the absolutely continuous invariant
density was calculated analytically in Ref.~\cite{Pel84}. Inspired by
this groundbreaking work and its cross-link to a simple diffusive
model \cite{RKla01b}, in Ref.~\cite{SaKl19} a spatially extended map
defined on the whole real line was introduced, where a suitably
adapted version of the Pelikan map generated long-time diffusion. At a
critical parameter value, determined by the Lyapunov exponent of this
model being zero, the invariant density of the associated Pelikan-type
map on the unit interval was computed numerically and found to be
non-normalisable, known as an infinite invariant density
\cite{Aar97,Zwei09,Kla13,AKB20}. Consequently, the dynamics of the
spatially extended system generated anomalous diffusion. Pelikan's
paper is also related to an important body of rigorous mathematical
works about random maps (see, e.g.,
Refs.~\cite{Blank01,HaYa23,Nis23,HaYa23} and further references
therein). Relevant for our setting are especially recent results by
Homburg, Kalle et al.: Starting from random logistic maps exhibiting
intermittency, synchronisation \cite{abbasi2018iterated} and a
respective transition in the invariant measure \cite{HKRVZ22} they
studied the decay of correlations in such systems \cite{KaZe23} as
well as invariant measures and associated Lyapunov exponents in
piecewise linear random maps that are more general than the Pelikan
map \cite{KaMa22,HoKa22}.

Motivated by this framework, in the present paper we explore the
non-trivial dynamical transition scenario of the original Pelikan map
in more depth. In particular, we calculate a suitably coarse-grained
functional form of the invariant density, a simple ergodicity breaking
parameter, and the position autocorrelation function under parameter
variation by characterising this transition in terms of these three
quantities. Exact and approximate analytical results are compared with
computer simulations. Our results for the invariant density enable us
to precisely identify the different dynamical regimes exhibited by
this random map during the transition. By considering a mean current
as an observable we obtain numerical evidence for ergodicity breaking
at the transition point. Correspondingly, we find that the
autocorrelation function displays a transition from exponential to
power-law decay when approaching the critical parameter value. We thus
arrive at a detailed characterisation of the whole transition scenario
in this random map. This transition, being of on-off type as in a
certain class of deterministic dynamical systems
\cite{Pik84,FuYa85,FuYa86,PiGr91,PST93,HPH94,OtSo94,HaMi97,HHF99,MHB01,SaKl19,HoRa20},
turns out to be very different from the seemingly similar transition
to intermittency in the paradigmatic Pomeau-Manneville map
\cite{PoMa80,Mann80,GaWa88,Wang89a,MeZw15}. What we study here is
probably one of the simplest non-trivial examples of a random
dynamical system. The great advantage of this system is that many
results can be proved analytically, demonstrating for a generic case
how anomalous properties of random dynamical systems emerge from first
principles. Many of the phenomena that we describe in this paper may
arise in other, more complex systems as well, as long as there is a
transition scenario where a parameter is changed that controls the
relative likelihood of a contracting dynamics alternating with an
expanding one \cite{SaKl19}.

We proceed as follows: In Sec.~\ref{sec-inv-den} we introduce our
random dynamical system, the Pelikan map, and calculate its
corresponding invariant density both analytically and numerially. We
summarise our results in a table outlining the whole transition
scenario. Furthermore, we provide numerical evidence for weak
ergodicity breaking by studying a relevant observable. In
Sec.~\ref{sec-acf} we calculate the position autocorrelation function
of this random map analytically exactly, approximately, and by
simulations, and compare the results with each
other. Section~\ref{sec-outlook} concludes with a brief summary and an
outlook. All detailed calculations are shifted to the Appendix.

\section{Transition to intermittency in the Pelikan map}\label{sec-inv-den}

\subsection{The Pelikan map}
We consider one-dimensional maps
\begin{equation}
  x_{n+1} = T(x_n) \:,
  \label{eq:1dmap}
\end{equation}
  where $n\in\mathbb{N}$ is discrete time and $x\in[0,1)$ the position
    of a point on the unit interval. Our maps are piecewise linear
    with uniform slope $s>0$.  To construct a simple random dynamical
    system $T(x)$, we sample from two maps $T_1(x),T_2(x)$ with
    different values of the slope $\{s_1,s_2\}$ randomly in time. That
    is, at each time step $n$ one of the two maps is chosen randomly
    with a probability $p\in[0,1]$, respectively $1-p$. In particular,
    we choose $s_1>1$ and $s_2<1$ so that $T_1(x)$ is expanding with
    positive Lyapunov exponent $\lambda_1 =\ln s_1>0$ and thus
    chaotic, while $T_2(x)$ is contracting with negative Lyapunov
    exponent $\lambda_2=\ln s_2<0$. More specifically, we consider the
    case of
\begin{equation}
T(x) = \begin{cases}
T_1(x)=s_1x \text{ mod }1 &\text{ with probability } p \\
T_2(x)=s_2x &\text{ with probability } 1-p\:.
\end{cases}
\label{rdsT}
\end{equation}
For the parameter values $s_1 = 2 = 1/s_2$ we call the random map $T$
the {\it Pelikan map}, as, to our knowledge, it was first studied by
Pelikan \cite{Pel84}, later also in Refs.~\cite{LOC90,AAN98}; see
Refs.~\cite{Blank01,maldonado2021phase,HoKa22} for related examples,
In the remainder of the paper, we mostly focus on the Pelikan map. The
values of $p=0$ and $p=1$ define two deterministic limiting cases for
which $T$ is well understood: In the latter case $T=T_1(x)$ reproduces
the well-known {doubling map} as studied in many textbooks
\cite{KaHa95,Ott,LaMa,Beck}. The former case $T=T_2(x)$ yields a
simple contraction on the unit interval, where all points converge
towards the stable fixed point at $x=0$. However, for $0 < p < 1$
stochastic randomness sets in generating a dynamical transition from
global contraction at $p = 0$ to uniform chaos at $p = 1$. To
understand this transition under variation of $p$ is the main topic of
our paper.

\subsection{Invariant density}

\begin{figure}
\centering
\subfloat[$p = 0.999$]{
	\includegraphics[width = 0.23\linewidth]{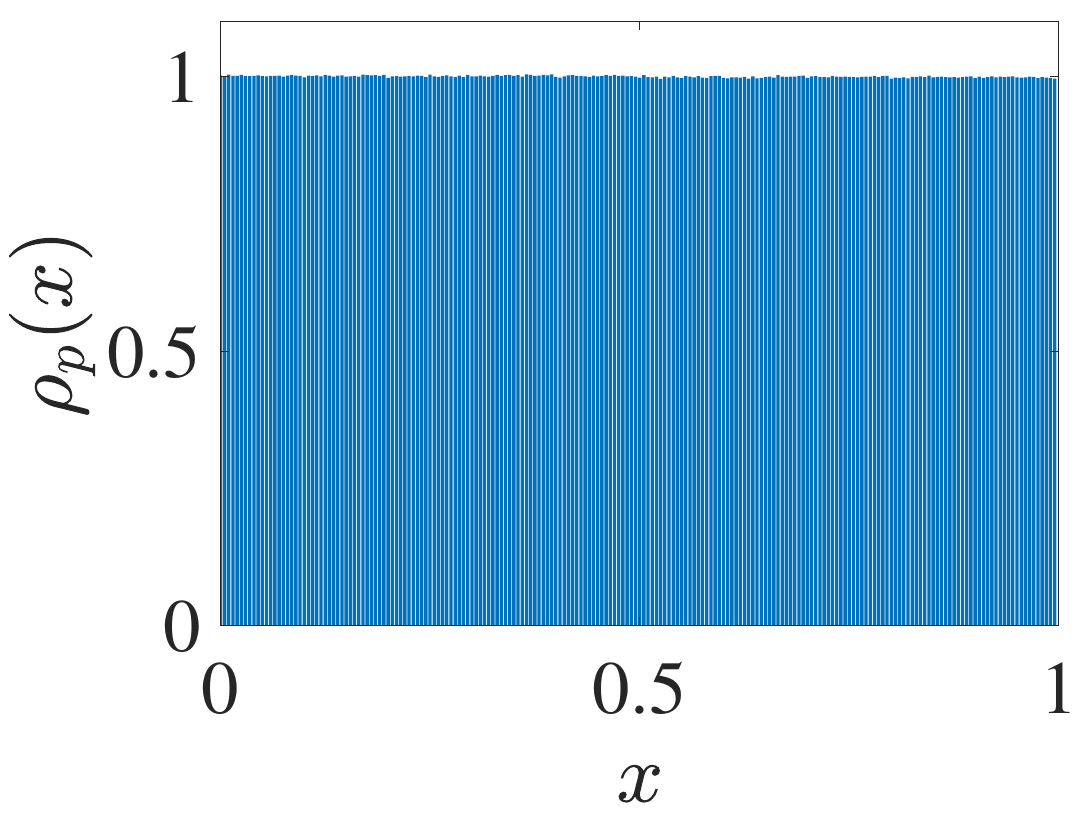}
	} 
\subfloat[$p = 0.8$]{
	\includegraphics[width = 0.23\linewidth]{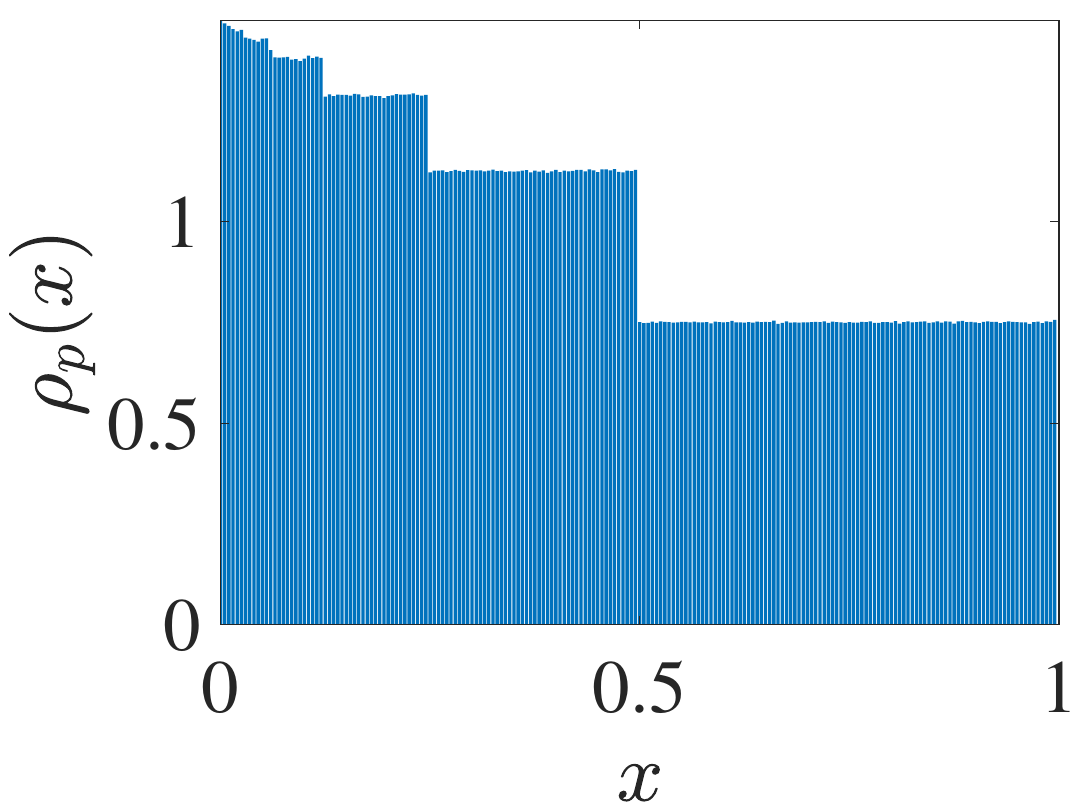}
	} 
\subfloat[$p = 0.65$]{
	\includegraphics[width = 0.23\linewidth]{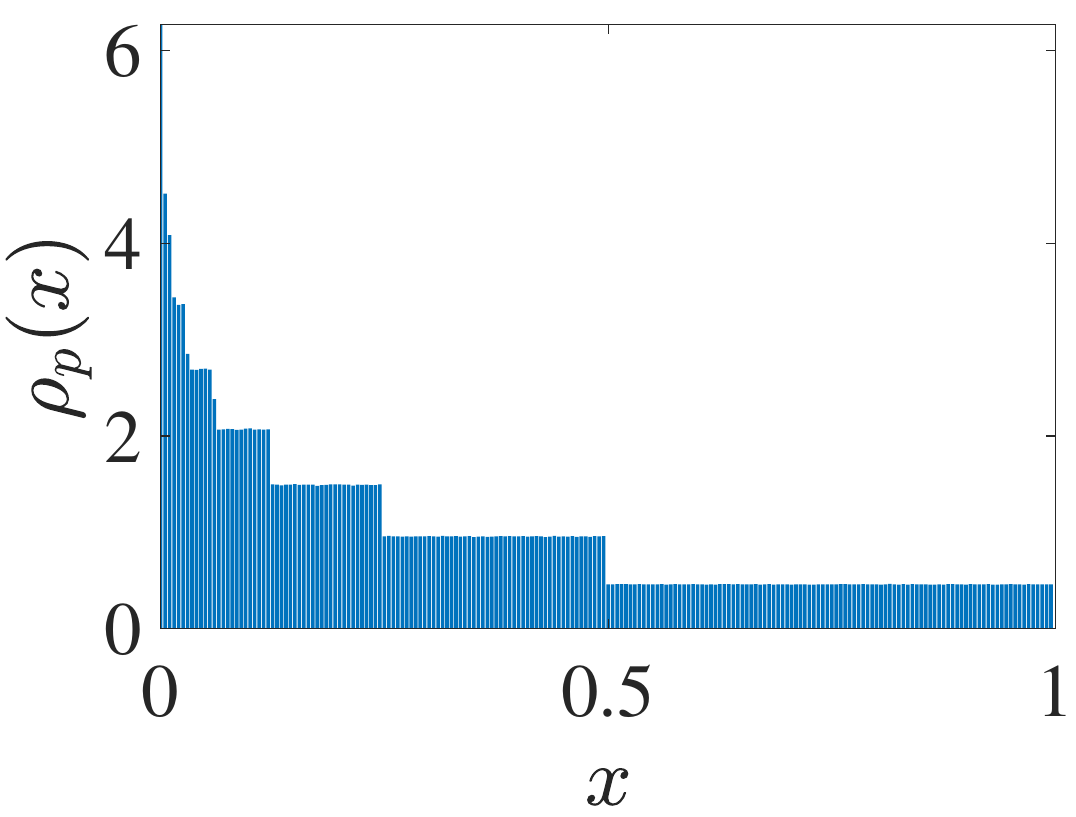}
	} 
\subfloat[$p = 0.501$]{
	\includegraphics[width = 0.23\linewidth]{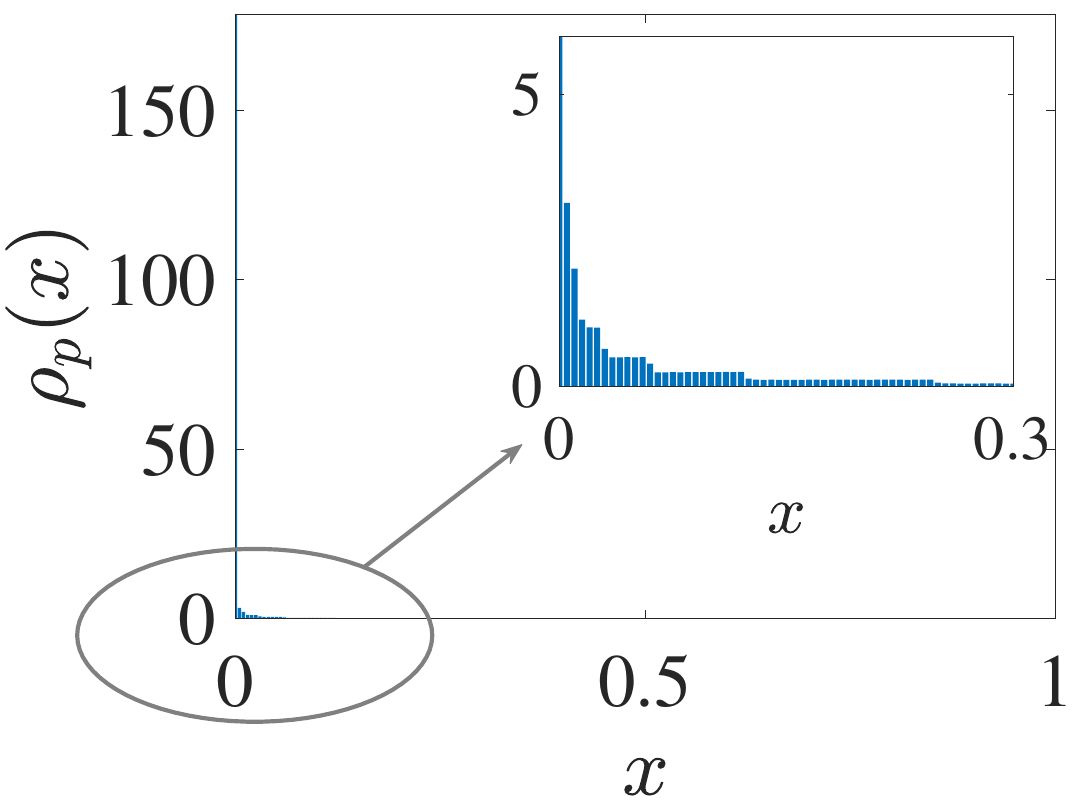}
	} 

\subfloat[\label{subfig: f}]{
	\includegraphics[width = 0.23\linewidth]{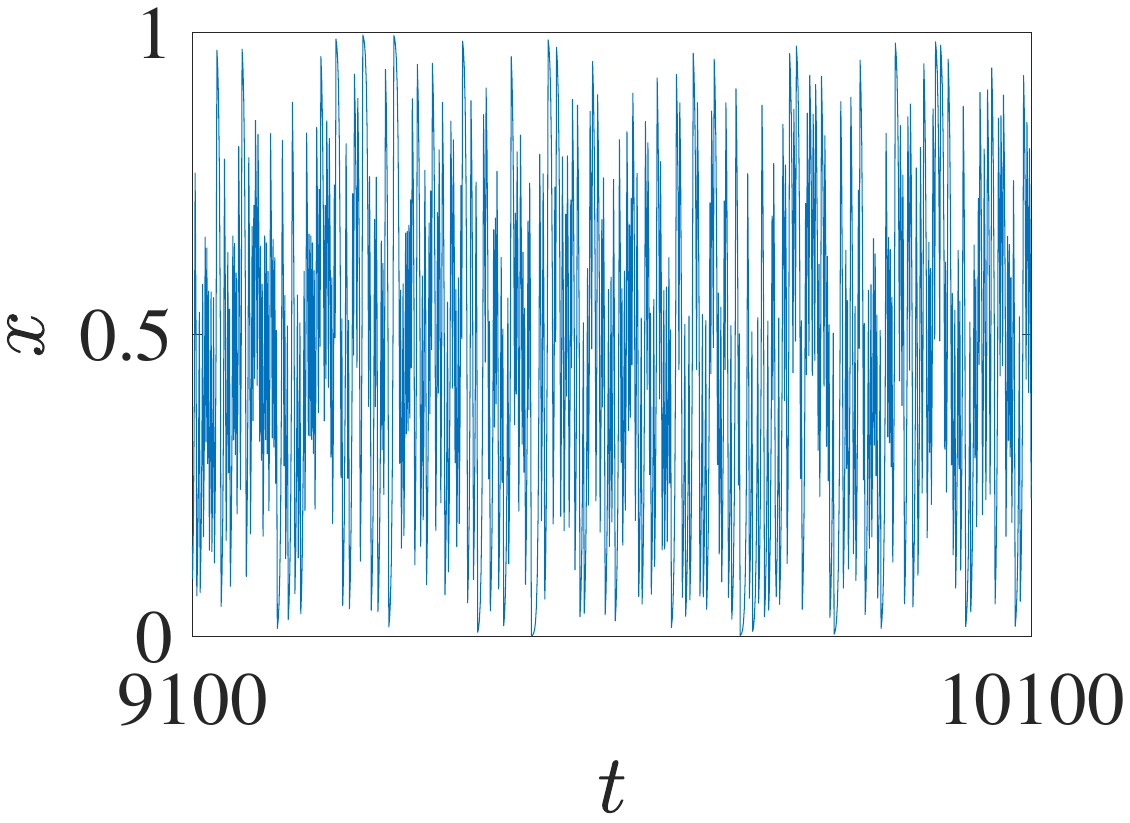}
	} 
\subfloat[\label{subfig: g}]{
	\includegraphics[width = 0.23\linewidth]{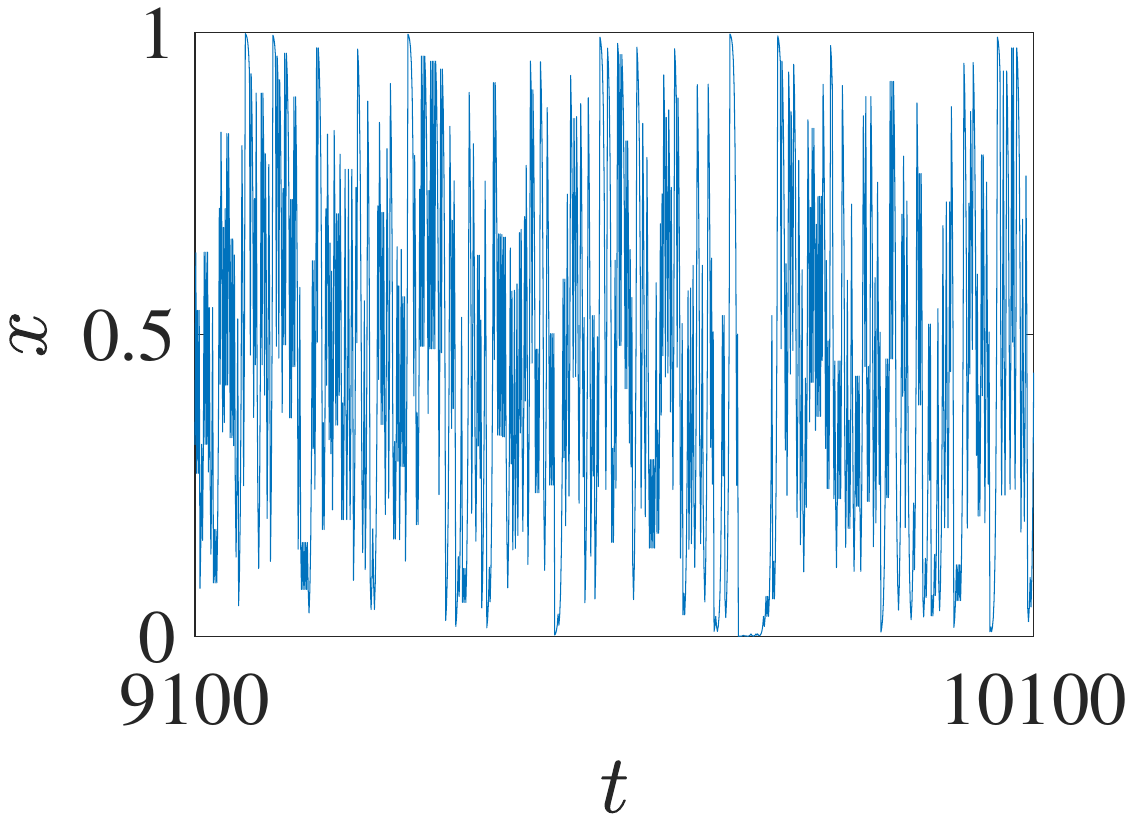}
	} 
\subfloat[\label{subfig: i}]{
	\includegraphics[width = 0.23\linewidth]{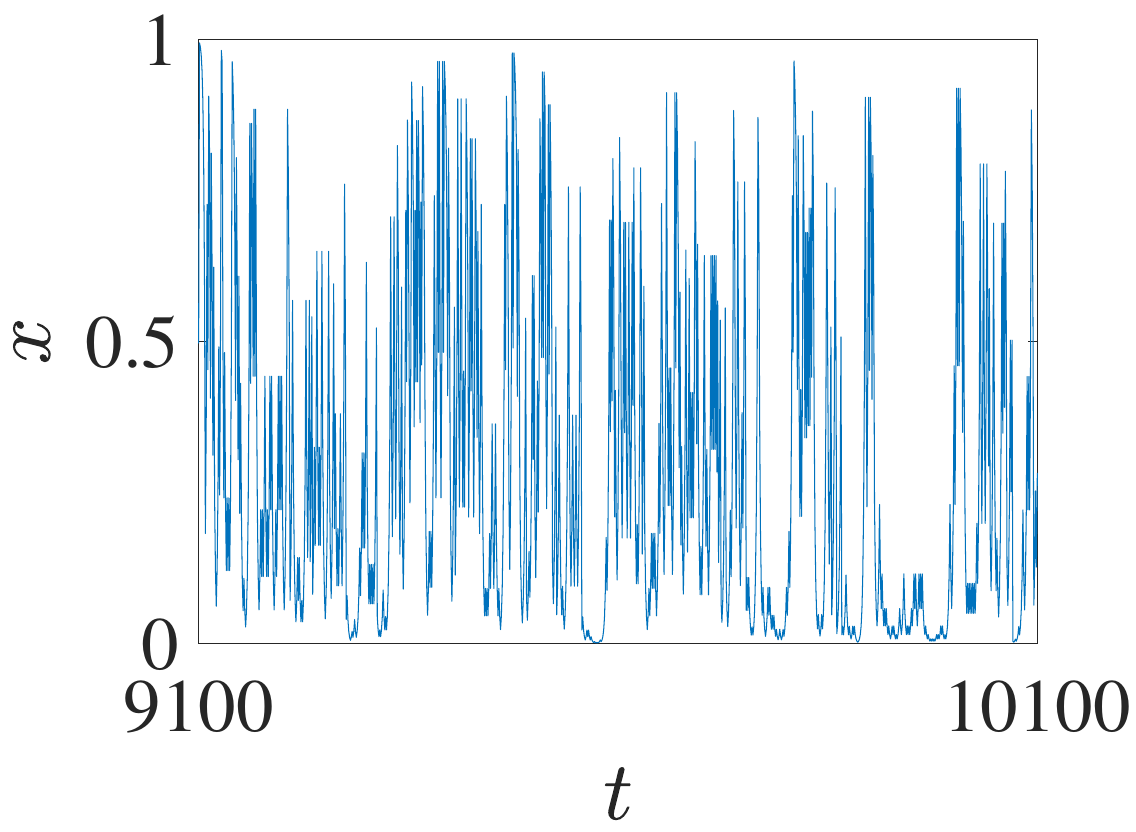}
	} 
\subfloat[\label{subfig: j}]{
	\includegraphics[width = 0.23\linewidth]{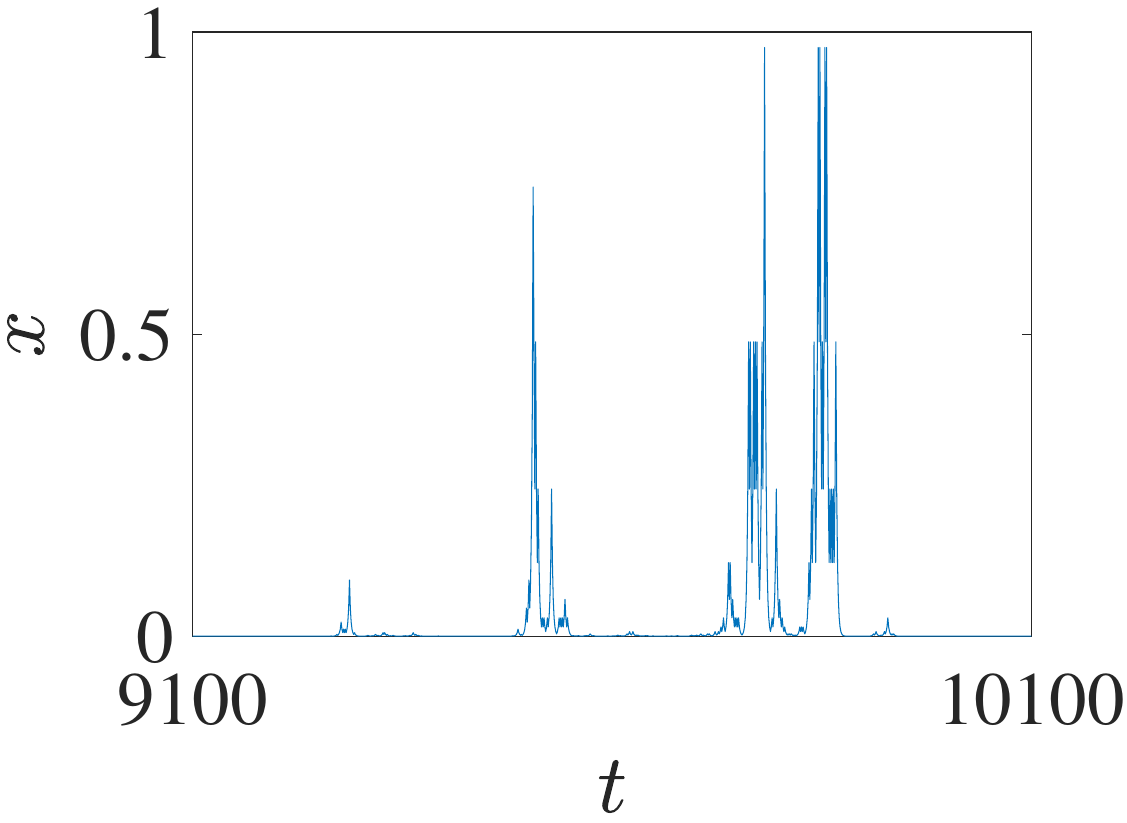}
	} 
\caption{Invariant density $\rho_p(x)$ (upper row, labels (a) to (d))
  for different values of the sampling probability $p \in (0.5, 1)$
  and corresponding typical trajectories (lower row, labels (e) to
  (h)) of the Pelikan map $T$, defined by Eq.~\eqref{rdsT} for
  particular values of the two slopes as explained in the text. The
  values of $p$ are given below the upper row. Each histogram is
  generated from computer simulations by iterating $10^4$ initial
  points (uniformly randomly chosen) for $10100$ time steps, with the
  first $100$ iterations discarded to eliminate transients. The bin
  size in each histogram is $\Delta x=1/200$. One can see that the
  density becomes singular around $x=0$ for $p\to1/2$, cf.\ the values
  along the $y$ axis and the blow-up in the inset of (d). The time
  window for all trajectories in the lower row is $t \in [9100,
    10100]$.}

\label{inv-den-plots}
\end{figure}

We first study the invariant density $\rho_p(x)$ of the Pelikan map
under variation of $p$, which characterises the equilibrium states of
the random dynamics. For $0 \leqslant p < 1/2$, $\rho_p(x)$ is
trivially the delta function at the stable fixed point $x = 0$, as the
map $T$ is on average contracting \cite{KaHa95}.  For $1/2 < p
\leqslant 1$ it has been proven in Ref.~\cite{Pel84} that $T$ exhibits
a unique absolutely continuous invariant measure, whose support is all
of $[0, 1)$, with an invariant density that is piecewise constant over
  the partition $\{ [ 2^{-(n+1)}, 2^{-n}] \}_{n \in \mathbb{N}_0}$ of
  the unit interval.  This is shown in the first row of
  Fig.~\ref{inv-den-plots}, which depicts $\rho_p(x)$ obtained from
  computer simulations for representative values of $p$. In the second
  row of Fig.~\ref{inv-den-plots}, we see that close to $p = 1/2$ a
  typical trajectory exhibits intermittency
  \cite{PoMa80,Mann80,Ott,LaMa}, that is, it displays irregular
  chaotic bursts within long regular motion close to the stable fixed
  point.

\begin{figure}[H] 
	\centering
\subfloat[]{
	\includegraphics[width = 0.36\linewidth]{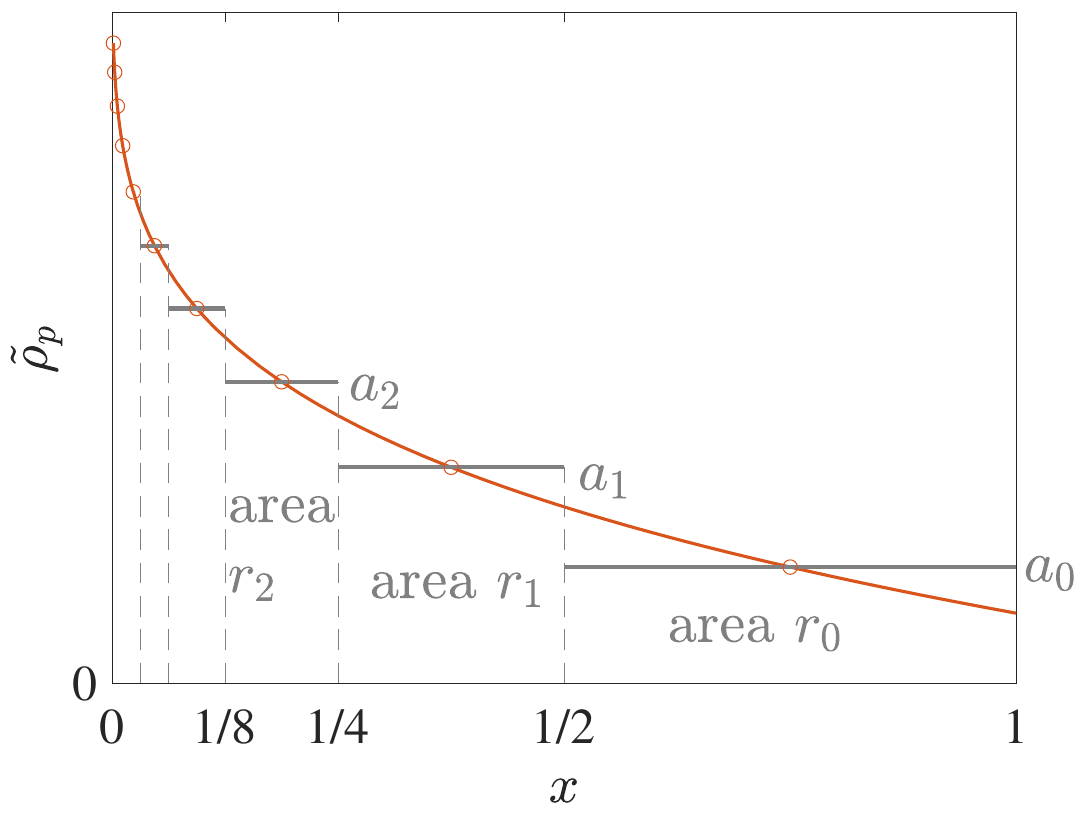}
	\label{inv-den-curve-illu}
	} 	
\subfloat[]{
	\includegraphics[width = 0.36\linewidth]{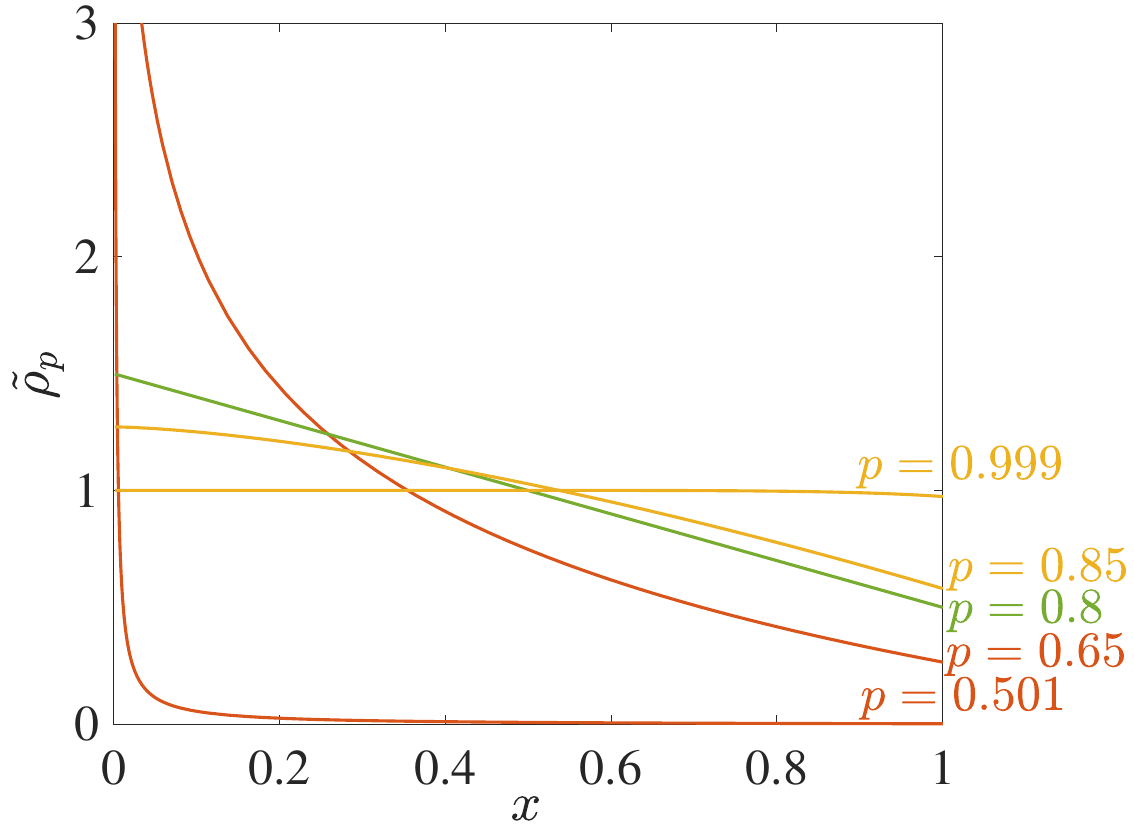}
	\label{inv-den-4curves}
	}
\caption{(a) Construction of the coarse-grained invariant density
  $\tilde{\rho}_p(x)$ (in orange, curved line) for the Pelikan map $T$
  with $p \in (0.5, 1)$. The original exact invariant density
  $\rho_p(x)$ is given by the piecewise constant amplitudes $a_i$ (in
  black) and the associated areas $r_i\:,\:i \in \mathbb{N}_0$,
  cf.\ Fig.~\ref{inv-den-plots}. The corresponding $\tilde{\rho}_p$ is
  defined by the mid-points of all amplitudes.  (b) Invariant density
  $\tilde{\rho}_p(x)$, plotted by using
  Eqs.~\eqref{eq:cgd1},\eqref{eq:cgd2}, for $p = 0.999$, $0.85$,
  $0.8$, $0.65$ and $0.501$ (cf.\ Fig.~\ref{inv-den-plots}). The
  change of colour indicates the change in convexity of the curves at
  $p = 0.8$, cf.\ also the $p$ values to the right.}
\end{figure}

Along with the existence and uniqueness result for $\rho_p(x)$, in
Ref.~\cite{Pel84} an iterative formula has been derived for it
yielding the value on each partition part where the density is
piecewise constant. While this formula is exact, it does not elucidate
the coarse grained, generic functional form of $\rho_p(x)$ and how it
changes under variation of $p$. As we will show below, calculating
such an expression makes it easier to characterise the whole
transition, and to compare $\rho_p(x)$ with the invariant density of
other intermittent chaotic dynamical systems, in particular the
Pomeau-Manneville map \cite{PoMa80,Mann80}. For this purpose, we
determine a coarse-grained invariant density by joining the midpoints
of all pieces where $\rho_p(x)=const.$ (cf.\ the orange curve in
Fig.~\ref{inv-den-curve-illu}). Its functional form can be derived
analytically from the normalisation condition for $\rho_p(x)\:,\:p \in
(1/2, 1]$. By using the amplitudes $a_i :=
\rho_p\rvert_{(2^{-(i+1)}, 2^{-i})}$ and the areas under constant
pieces of the invariant density $\rho_p$, $r_i :=
\int_{2^{-(i+1)}}^{2^{-i}} \rho_p dx\:,\:i \in \mathbb{N}_0$,
cf.\ Fig.~\ref{inv-den-curve-illu}, the normalisation reads $\sum_{i =
  0}^{\infty} r_i = 1$. Furthermore, $a_i$ and $r_i$ satisfy $a_n =
2^{n+1}r_n$ and the following recursion relations \cite{Pel84},
\begin{equation}
\frac{r_{n+1}}{r_n} = \frac{-1 + \left( \frac{2(1-p)}{p}\right)^{n+2}}{-2 + 2\left( \frac{2(1-p)}{p}\right)^{n+1}}, \quad \frac{a_{n+1}}{a_n} = 2\frac{r_{n+1}}{r_n}, \quad n \in \mathbb{N}_0.
\label{relation-r-a}
\end{equation}
This gives for the first amplitude $a_0 = (2p-1)/p$, and
recursively we have (see App.~\ref{appendix-inv-den-curve})
\begin{equation}
a_n = \frac{-1 + \left( \frac{2(1-p)}{p}\right)^{n+1}}{-1 + \frac{2(1-p)}{p}}a_0 = \frac{2p-1}{3p-2}\left[ 1 - \left( \frac{2(1-p)}{p}\right)^{n+1}\right].
\label{an}
\end{equation}
To distinguish the exact discontinuous invariant density $\rho_p$ from
its coarse-grained functional form, we denote the latter by
$\tilde{\rho}_p$.  This is in turn given by taking the limit $n \to
\infty$ of the mid-points $(3/2^{n+2}, a_n)$
leading to (see again App.~\ref{appendix-inv-den-curve}),
\begin{equation}
\tilde{\rho}_p(x) = A (1-Bx^{-1+C}), \quad p \in \left( \frac{1}{2}, 1\right), p \neq \frac{2}{3}, \label{eq:cgd1}
\end{equation}
where 
\begin{equation}
A(p) = \frac{2p-1}{3p-2}, \quad B(p) = \left( \frac{2(1-p)}{p}\right)^{\frac{\ln 3}{\ln 2} - 1}, \quad C(p) = \frac{1}{\ln 2}\ln \frac{p}{1-p}. \label{eq:cgd2}
\end{equation} 
The invariant density curves $\tilde{\rho}_p$ for various values of $p
\in (0.5, 1)$ are plotted in Fig.~\ref{inv-den-4curves}. For $p \to
1^-$ we see that $\tilde{\rho}_p \to 1$, which is the result for the
deterministic {doubling map} \cite{KaHa95,Ott,LaMa,Beck}. However,
as $p \to 1/2^+$ we have $C(p) \to 0$, consequently
$\tilde{\rho}_p(x)$ behaves as $1/x$ (normalisation is governed by the
factor $A$). In particular, at $p = 4/5$ we have $A = 3/2$, $B = 2/3$,
$C = 2$ and the curve becomes linear, $\tilde{\rho}_{p = 0.8}(x) = 3/2
- x$.  Consequently, as the parameter $p$ decreases from $1$ to $0$,
the random map $T$ undergoes a rather non-trivial dynamical
transition: At $p=1$ we have a uniform density $\tilde{\rho}_p$
representing fully chaotic dynamics. For $1>p$ this density becomes
non-uniform by decreasing monotonically. At $p=4/5$ the curvature of
the density changes from concave to convex. At $p=2/3$ the density
develops a singularity at $x=0$, which remains normalisable for $2/3 >
p > 1/2$ (see Ref.~\cite{Blank01} for related results regarding a
randomised tent map), hence the system exhibits stationary
intermittency. At $p=1/2$ there is a transition to non-stationary
intermittency, where the system possesses both an unbounded but
normalisable (in that sense `physical') invariant density $\delta(0)$
and an unbounded, non-normalisable (infinite) invariant density
$\tilde{\rho}_p$ (see Ref.~\cite{HaYa23} for a rigorous mathematical
analysis of this case).  Finally, for $1/2>p\geqslant 0$ global
contraction leads to a density that collapses onto the singular but
normalisable delta function at the fixed point $x=0$. A similar
transition scenario has been described for a related but more
complicated random map in
Ref.~\cite{maldonado2021phase}. {Intuitively, the transition from
  stationary to non-stationary dynamics at $p=1/2$ can be understood
  by considering the new variable $w_n=\log(x_n)$
  \cite{Pik84,PiGr91,LOC90,YOC91,Pik92,HPH94,AAN98}. The dynamics of
  this variable is then defined on the half-line $(-\infty,0]$ with
zero flux at $0$. Only for $p=1/2$ the random walk generated by $w_n$
is unbiased corresponding to a non-stationary density
\cite{Pik84,LOC90,YOC91,AAN98}.}

This rich scenario represented by $\tilde{\rho}_p$ under variation of
$p$ may be compared with the transition to intermittency exhibited by
the Pomeau-Manneville map, $P(x)=x+ sx^z \mod 1\:,\: s\geqslant
1\:,\:z\geqslant 1$, by varying $z$ for fixed $s$
\cite{PoMa80,Mann80,GaWa88,Wang89a,MeZw15}. At first view there
are some striking similarities: In analogy to the Pelikan map $T$, for
$s=1$ and $z=1$ this map $P$ reduces to the {doubling map}. And
similar to the density of $T$ for $p\to1/2$, for $z\to2$ the density
of $P$ becomes non-integrable. At the transition points both maps
develop infinite invariant densities \cite{Aar97,Kla13} with even the
same singularity, i.e., $\rho_{PM}(x)\sim 1/x$ for $P$ at $z=2$
\cite{Tha83} and $\tilde{\rho}_p\sim1/x$ for $T$ at $p=1/2$
\cite{PiGr91,FuYa85,HPH94,HaMi97,AAN98,HHF99,MHB01,SaKl19}. However,
in further detail these two transition scenarios are completely
different: $P$ remains deterministic for all $z$ and is characterised
by a smooth invariant density $\rho_{PM}(x)\sim x^{1-z}$ for all $z>1$
\cite{Tha80,Tha83,Zwei98,Tha00,KoBa10,AkBa13} while $T$ is random for all
$0<p<1$, and its exact invariant density $\rho_p$ is piecewise
constant, see Eqs.~\eqref{relation-r-a},\eqref{an}.  Furthermore, in
sharp contrast to the coarse-grained density $\tilde{\rho}_p$ of $T$,
see Eqs.~\eqref{eq:cgd1},\eqref{eq:cgd2} and as discussed above, the
transition scenario of $P$ is much simpler: Its density immediately
becomes singular for all $z>1$ and is still integrable for $1<z<2$ but
as a convex power law. It is indeed known \cite{LOC90,AAN98,SaKl19}
that the transition exhibited by $T$ belongs to the class of on-off
intermittency
\cite{Pik84,FuYa85,FuYa86,PiGr91,PST93,HPH94,OtSo94,HaMi97,HHF99,MHB01,HoRa20},
which is different to the one of $P$.  We remark that the Pelikan map
appears to be one of the very few examples of a random dynamical
system for which the invariant density can be calculated exactly
analytically for all parameter values.

\subsection{Dynamical instability}
We now discuss dynamical stability properties of the Pelikan map in
terms of its Lyapunov exponent $\lambda$ under variation of $p$. In
order to do so, we first refine our notation for capturing particular
realisations of a noise sequence. Let $\omega_i$ be a random variable
for which at any time step $i\in\mathbb{N}$ one of two integer values
$\omega_i\in\{1,2\}$ is chosen randomly with probability $p$,
respectively $1-p$. These two values determine correspondingly which
of the two maps $T_i(x)\:,\:i={1,2}$, is applied at time step $i$,
according to Eq.~\eqref{rdsT}, for generating the position at the next
time step $i+1$, according to Eq.~\eqref{eq:1dmap}. Let
$\Omega_n=\{\omega_i\}_{i=1}^n$ denote the realisation of a particular
noise sequence up to time $n$. With the symbol $\Omega$ we label the
dependence of an observable on the realisation of a particular noise
sequence $\Omega_n$ in the limit of $n\to\infty$. Based on these
preliminaries, we define two different types of averages for the
Pelikan map $T$ \cite{CSNG21}

\begin{enumerate}
\item The time-averaged Lyapunov exponent
  \cite{chekroun2011stochastic,SaKl19}
\begin{equation}
  \lambda_t=\int \lambda_t(\Omega)\,\varrho(\Omega) \,d\Omega
\end{equation}
with
\begin{equation}
  \lambda_t(\Omega) = \lim_{n \to \infty} \frac{1}{n}\sum_{i = 1}^n\ln |T'(x_i(\Omega_i))|\:,
\end{equation}
where $\varrho(\Omega)$ is the distribution of the noise. With
$x_i(\Omega_i)$ we denote the dependence of the position of a point at
time step $i$ on the particular realisation of the noise sequence
$\Omega_i$. In the deterministic case, i.e., $p=0$ or $p=1$, this
average boils down to
\begin{equation}
  \lambda_t^d = \lim_{n \to \infty} \frac{1}{n}\sum_{i = 1}^n\ln |T'(x_i)|\:.
\end{equation}

\item The ensemble-averaged Lyapunov exponent 
\begin{equation}
  \lambda_e = \int \ln |T'(x(\Omega))| \,\rho_p(x(\Omega),\Omega) \,d\Omega \, dx \:,
  \label{eq:le}
\end{equation}
where $\rho_p(x)=\int\rho_p(x(\Omega),\Omega)d\Omega$ holds for the
invariant density of the system, by assuming that it exists. In the
deterministic case this average boils down to
\begin{equation}
  \lambda_e^d = \int \ln |T'(x)| \,\rho_p(x) dx \:.
\end{equation}

\end{enumerate}
Let us now calculate both types of Lyapunov exponents for map $T$. Due
to the uniform slope of $T$, the time-averaged Lyapunov exponent is
obtained straightforwardly to \cite{SaKl19}
\begin{equation}
  \lambda_{t} = p\ln 2 + (1-p)\ln \frac{1}{2}=(2p-1)\ln 2\:.
\end{equation}
It is thus well-defined for all $p$. In particular, it vanishes at $p
= 1/2$, which coincides with the transition to non-normalisability of
the invariant density \cite{SaKl19}.  {We remark that in
  Refs.~\cite{Pik84,PiGr91,Pik92,OtSo94,HPH94,AAN98} this transition
  has been analysed for a much wider class of random dynamical
  systems.}  For the ensemble-averaged Lyapunov exponent we observe
that Prob$\left[\ln |T'(x(\omega_i))| = \ln 2\right] = p$ and
Prob$\left[\ln |T'(x(\omega_i))| = \ln (1/2)\right] = 1-p$ at any time
step $i$, according to Eq.~\eqref{rdsT}, and due again to the
uniformity of the slope therein. Averaging over the (in this case
dichotomic) noise in Eq.~\eqref{eq:le} then yields
\begin{equation}
\lambda_e = p\ln 2 + (1-p)\ln \frac{1}{2}=(2p-1)\ln 2\:,\:p\neq \frac{1}{2}\:.
\end{equation}
At $p=1/2$ we have the infinite density $\rho_p$, whose
non-normalisability defies a unique result for $\lambda_e$. We
conclude that time- and ensemble-averaged Lyapunov exponent for map
$T$ are equal, $\lambda_t=\lambda_e$, for $p \neq 1/2$. In this sense,
we consider the system to be ergodic for $p \neq 1/2$, and clearly it
is chaotic when $1/2 < p \leqslant 1$.

It should be noted that, in some random dynamical systems, the
numerically observed Lyapunov exponent can be negative yet the system
still exhibits chaotic behaviour \cite{doan2018hopf,SDLR19}. In such
systems the Lyapunov exponent has no definite value and is described
by a distribution with a non-zero tail in the positive regime,
although the peak may lie in the negative regime. For more general
random dynamical systems, Lyapunov exponents may not reveal dynamical
properties such as topological bifurcations of attractors
\cite{SDLR19}. Along these lines, for the Pelikan map with $0 < p <
1/2$ one can observe chaotic dynamics for finite time, in the sense of
a chaotic transient, with a positive probability, but the Lyapunov
exponent is negative.  In contrast, the Pomeau-Manneville map is
characterised by ergodicity and a positive Lyapunov exponent for
$1\leqslant z\leqslant 2$, while it exhibits a stretched exponential
dynamical instability leading to a zero Lyapunov exponent, and
infinite ergodicity, for $z>2$ \cite{GaWa88,Wang89a,Kla13}.  In
Table~\ref{table:dyn} we summarise the transition to intermittency in
the Pelikan map.

\begin{center}
  \begin{table}
\begin{tabular}{| m{1.6cm} | m{5.4cm} | m{4.8cm} | m{5cm} |}
\hline
parameter & invariant density $\rho_p(x)$, $x \in [0, 1]$ & Lyapunov exponent & dynamics\\
\hline \hline \vspace{1em}
$p = 1$ & uniform density $\rho_1(x) = 1$ & $\lambda_t^d = \lambda_e^d = \ln 2$ & $T = T_1$ uniformly chaotic\\
\hline \vspace{1em}
$1>p \geqslant\frac{2}{3}$ & bounded and normalisable & $\lambda_t = \lambda_e = (2p-1)\ln 2$ & chaotic\\
\hline \vspace{1em}
$\frac{2}{3}>p>\frac{1}{2}$ & unbounded but normalisable & $\lambda_t = \lambda_e = (2p-1)\ln 2$ & stationary intermittency\\
\hline \vspace{1em}
$p = \frac{1}{2}$ & unbounded and non-normalisable & $\lambda_t = 0$, $\lambda_e$: undefined & non-stationary intermittency\\
\hline \vspace{1em}
$\frac{1}{2}>p>0$ & $\delta(0)$ & $\lambda_t = \lambda_e = (2p-1)\ln2$ & global contraction on average\\
\hline \vspace{1em}
$p = 0$ & $\rho_0(x)=\delta(0)$ & $\lambda_t^d = \lambda_e^d = -\ln 2$ & $T = T_2$ global contraction\\

\hline
\end{tabular}
\caption{Summary of dynamical properties of the Pelikan map,
  characterised by the invariant density and the  Lyapunov
  exponent.}
\label{table:dyn}
\end{table}
\end{center} 

We remark that one can generalise our above analysis to the case where
$s_1 = 1/s_2$ \cite{Yan21}. It turns out that for these specific
slopes the random dynamical system Eq.~\eqref{rdsT} still possesses a
regular partition with a piecewise constant invariant density. On this
basis one can show that the normalisability still changes at $p =
1/2$, always coinciding with the vanishing of the time-averaged
Lyapunov exponent given by $\lambda_t = (2p-1)\ln s_1$, and the
boundedness changes at $p = s_1/(s_1+1)$. The convexity change occurs
at $p = s_1^2/(s_1^2 + 1)$, as is shown in
App.~\ref{appendix-convexity}. We note that related, rigorous results
have been obtained in Ref.~\cite{Hom23}. Here, for $s_1,1/s_2\ge2$ it
was proven that in the corresponding random map there exist different
invariant measures under variation of $p$ depending on the associated
Lyapunov exponent, which supports our findings. A generalised version
of the setting in Ref.~\cite{Hom23} is studied in Ref.~\cite{HoKa22}.

\subsection{Ergodic properties}

\begin{figure}[H] 
	\centering
\subfloat[linear scale]{\includegraphics[width = 0.36\linewidth]{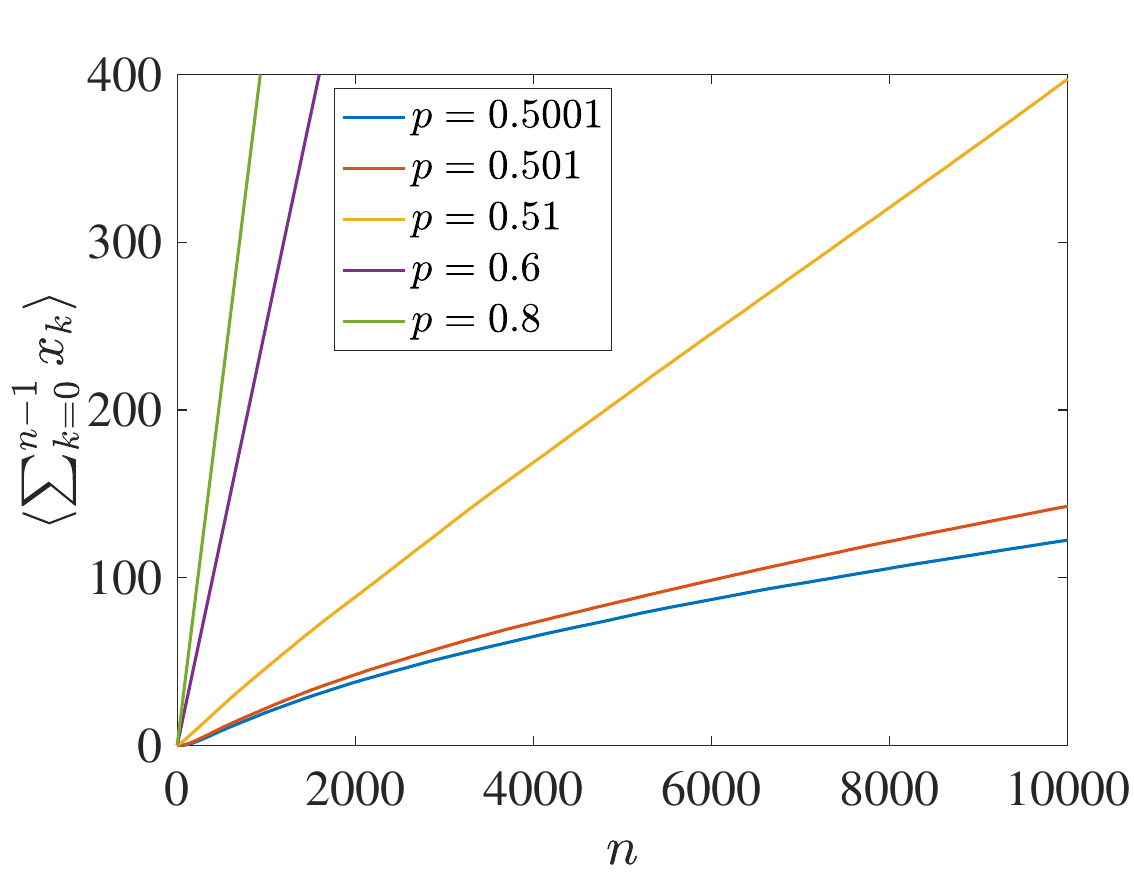}
}
\subfloat[double-logarithmic scale]{\includegraphics[width = 0.36\linewidth]{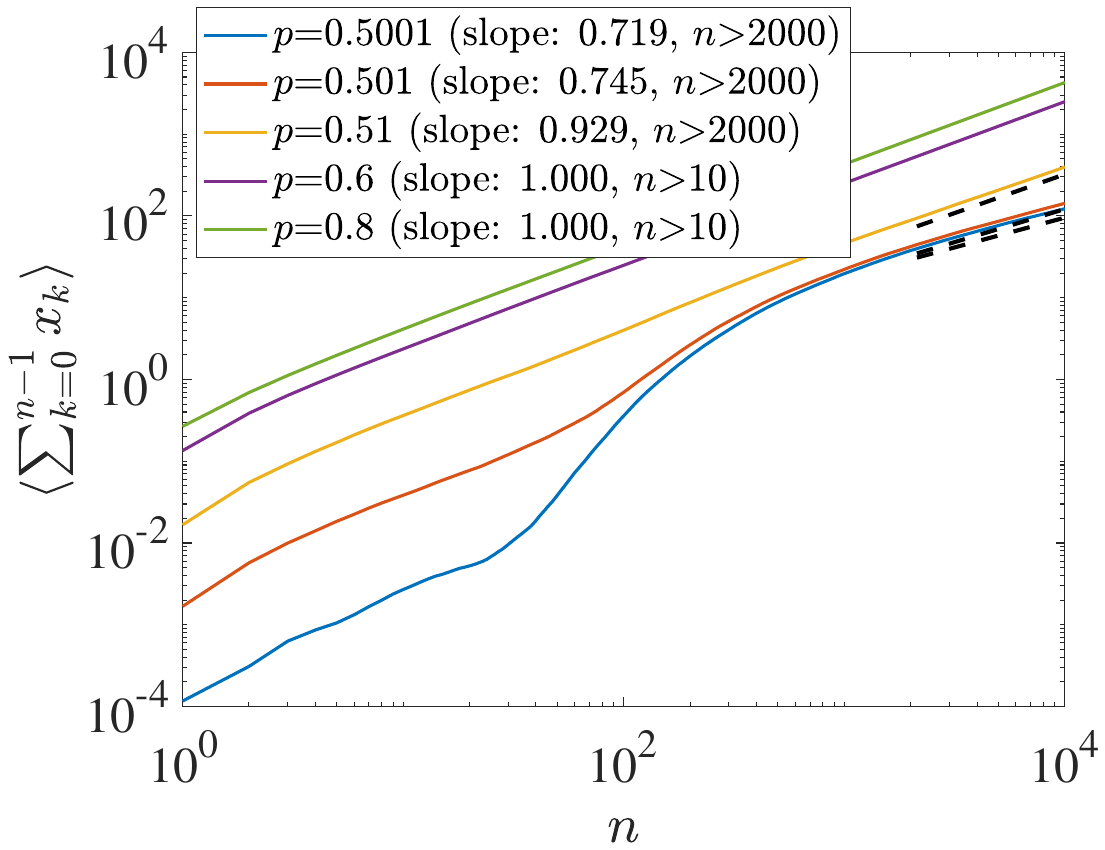}
} 
\caption{The noise-ensemble average $\langle \sum_{k = 0}^{n-1} x_k
  \rangle$ as a function of time $n$, taken over $10^4$ initial points
  (sampled directly from the invariant density formula with a
  truncation of $20$ subintervals on $[0, 1]$), and time up to $n =
  10^4$. (a) shows results on a linear scale, (b) depicts a
  double-logarithmic plot.  In (b), the dashed lines for $p = 0.5001,
  0.501$ and $0.51$ are linear fits from time $n > 2000$. The
  slopes are given in the legend, in comparison to $p = 0.6$ and
  $0.8$ (linear fits from $n > 10$), where the slopes are $1$.}
\label{aver-y}
\end{figure}

We finally study in some more detail the ergodic properties of the
Pelikan map with respect to the non-trivial transition at $p=1/2$. If
we consider a new variable $y_n$, defined via the Birkhoff sum of
iterates of the Pelikan map with the mean subtracted
\cite{beck1996dynamical},
\begin{equation}
y_{n+1} = y_n + x_{n-1} - \langle x\rangle = \sum_{k=0}^{n-1} (x_k - \langle x\rangle)\:,
\label{eq-yn}
\end{equation}
the average of $y_{n+1}$ can be determined by averages of $x_k$
and $x$ as
\begin{equation}
\langle y_{n+1}\rangle = \langle \sum_{k = 0}^{n-1} (x_k - \langle x \rangle) \rangle = \langle \sum_{k = 0}^{n-1} x_k \rangle - n\langle x \rangle \:.
\end{equation}
Here we denote by $\langle \ldots \rangle$ the combined noise and
ensemble average.  In Fig.~\ref{aver-y} we see from simulations that
for all $p>1/2$ the averaged Birkhoff sum $\langle \sum_{k = 0}^{n-1}
x_k \rangle$ increases linearly with time $n$, while for $p \to 1/2^+$
it changes to sub-linearly.  This nonlinear behaviour of $\langle
\sum_{k = 0}^{n-1} x_k \rangle$ can be considered as an indicator of
weak ergodicity breaking \cite{Bou92,BeBa05,Metz15}, in the sense that
$\langle y_{n+1} \rangle = n \left(\frac{1}{n}\langle\sum_{k=0}^{n-1}
x_k\rangle - \langle x \rangle \right) \neq0$ implies that the time
average of $x_k$ is not equal to the ensemble average for the Pelikan
map.

\section{Position autocorrelation function}\label{sec-acf}

Correlation functions are important quantities characterising
dynamical systems. Their decay in time is intimately related to
fundamental ergodic properties like mixing \cite{LaMa,Beck}. In the
form of velocity autocorrelation functions they can be used to
calculate transport coefficients of dynamical systems via
(Taylor-)Green-Kubo formulas \cite{KCKSG06,KKCSG06,Kla06,Do99}.  For
simple deterministic piecewise linear maps, including the {doubling map},
position autocorrelation functions have been calculated
analytically in Ref.~\cite{MSO81}; see also
\cite{yan2020distinguished}. Here we focus on the position
autocorrelation function of the Pelikan map, defined as the ensemble
average of the product of two positions at different times,
\begin{equation}
  \langle x_{k+t}x_t\rangle = \langle x_k x_0 \rangle = \int
  \rho_p(x_0) \,  T^k (x_0) x_0 \,dx_0\:,\: k,t \in \mathbb{N}\:.
  \label{eq:acf}
\end{equation}
Note that this ensemble average is defined with respect to the
invariant density $\rho_p$ of the map, according to which the initial
conditions are distributed. This implies that we can simplify the
first product to the second one by using stationarity, respectively
time translational invariance. In the following we first compute this
autocorrelation function numerically over the whole range of $p \in
\left( 1/2,1\right]$, up to somewhat larger times. We then calculate
  it exactly analytically for the same range of $p$ at the first three
  time steps. As calculating exact results for larger times becomes
  too tedious, we then develop an analytical approximation that gives
  some insight into the asymptotic decay particularly for $p\to1/2$
  and $p\to1$.

\subsection{Simulation results}

\begin{figure}[H]
	\centering
	\subfloat[semi-log scale in $k$ and nCF]{
	\includegraphics[width = 0.36\linewidth]{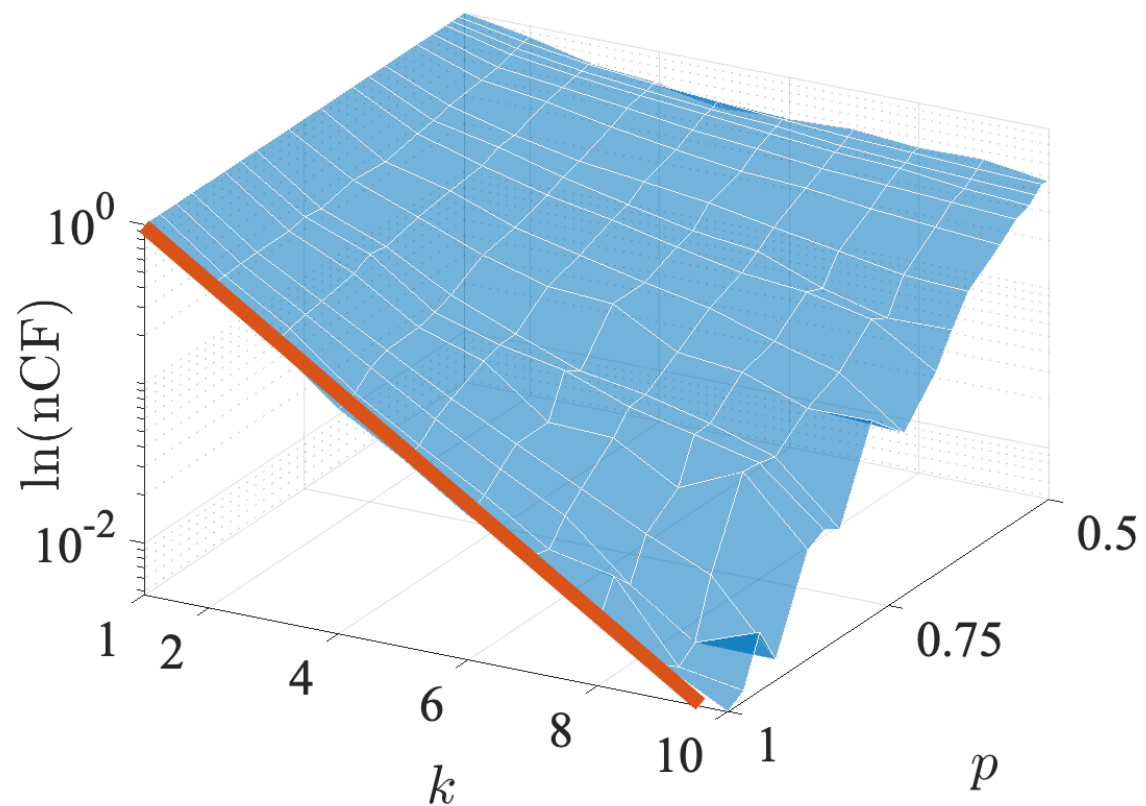}
	} 	
	\subfloat[log-log scale in $k$ and nCF]{
	\includegraphics[width = 0.36\linewidth]{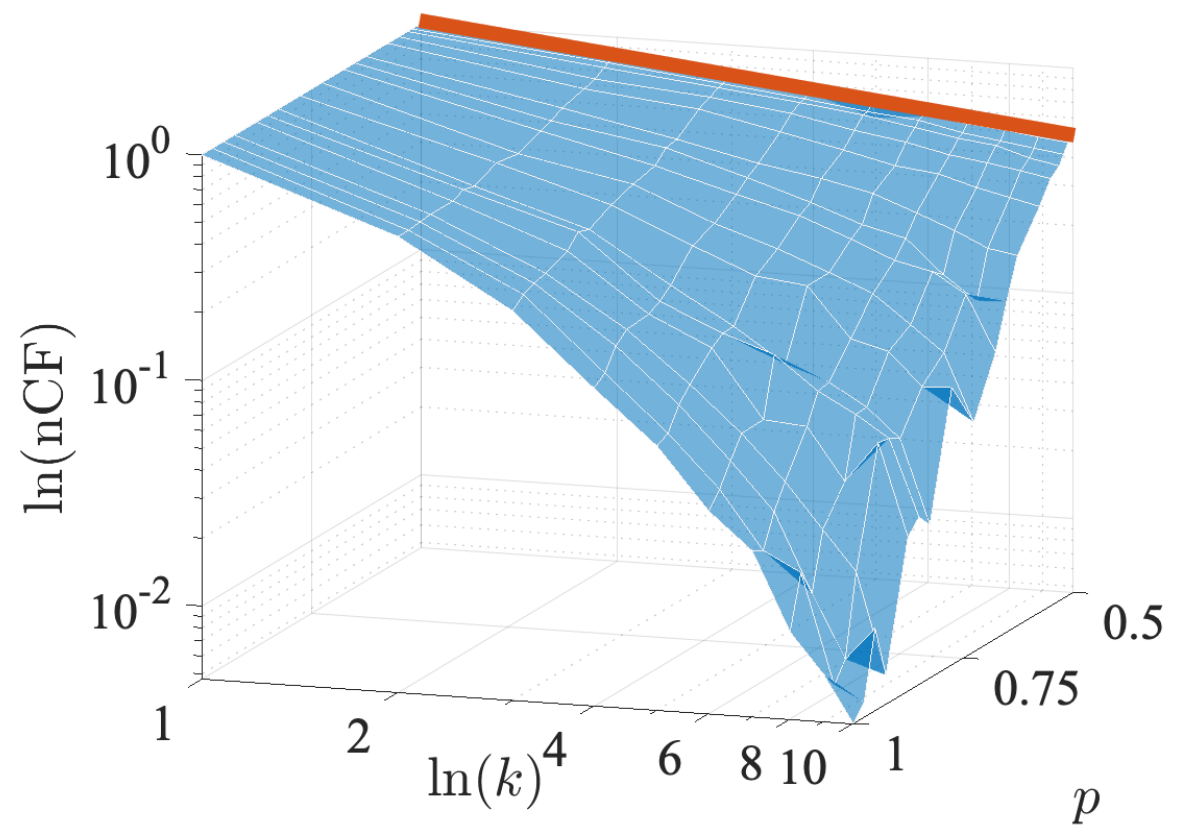}
	} 
\caption{Simulation results for the normalised position
  autocorrelation function nCF Eq.~\eqref{nCF-eqn} as a function of
  both $p \in (0.5, 1)$ and time $k = 1, ..., 9$. The graphs are
  generated by averaging over $10^5$ initial points sampled directly
  from the invariant density formula (with a truncation of $20$
  subintervals of the unit interval). $25$ values of $p$ have been
  used between $0.5001$ and $0.9999$, with more values closer to
  $0.5$. }
\label{nCFplot}
\end{figure}

To compare results of the position autocorrelation function
Eq.~\eqref{eq:acf} for different values of $p \in \left( 1/2,
1\right]$, we define the normalised correlation function (nCF)
\begin{equation}
  \text{nCF}(p, k) := \frac{\langle x_k x_0 \rangle - \langle x \rangle^2}{\langle x_0^2 \rangle - \langle x_0\rangle^2}\:,
\label{nCF-eqn}
\end{equation}
which we have computed numerically. There are two main subtleties for
the numerical simulations. First, by definition the Pelikan map
involves a binary shift operation (and its contraction
counterpart). But since any computer stores numbers in finite 0-1
bits, high-precision simulations are needed to accurately obtain
autocorrelations for large times, otherwise the computer memory would
run out of bits, and nothing can be shifted anymore. For our
simulations we have thus used a high-precision algorithm from the GNU
MPFR library in C \cite{FHLPZ07}. Second, as $p$ approaches the
critical value of $1/2$, where the invariant density becomes
non-normalisable, preparing such a highly singular initial
distribution becomes increasingly difficult. Instead of obtaining
initial conditions directly from the density formula, one could start
from an arbitrary distribution and wait until it approximately reaches
an equilibrium. However, for $p\to1/2$ the transient time becomes
extremely long. We have tested both methods but eventually decided to
use the former. Figure~\ref{nCFplot} presents simulation results for
Eq.~\eqref{nCF-eqn}.  In Fig.~\ref{nCFplot}(a) we see that, as
expected, there is clean exponential decay for $p=1$. That there is
exponential decay of correlations for $p>1/2$ in the limit of large
enough times was already proven in Refs.~\cite{Pel84,Blank01}, and our
numerical findings are in line with this result. Interestingly,
Fig.~\ref{nCFplot}(b) shows a transition to a clean power law decay
for $p\to1/2$.  {One might speculate whether the transition to
  unboundedness of the density at $p=2/3$ has an impact on the
  correlation decay, at least for shorter times. The surface in
  Fig.~\ref{nCFplot} indeed seems to flatten out a bit with respect to
  varying $p$ below $p=2/3$, but our simulations are not entirely
  conclusive in this respect. Generally, the transition between
  exponential and power law decay appears to be smooth, in the sense
  that a power law decay develops for smaller times by decreasing $p$,
  with eventually crossing over to exponential decay for larger times
  for all $p>1/2$, with the cross-over point moving to larger times
  for $p\to1/2$. Notably, this implies that there is not just
  anomalous dynamics exactly at the transition point of $p=1/2$,
  rather, the transition is marked by longer and longer anomalous
  transients when approaching the transition. This observation is in
  line with the time-dependence of the average Birkhoff sum plotted in
  Fig.~\ref{aver-y} showing long transients and corresponding findings
  for diffusion generated by a Pelikan-like map \cite{SaKl19}. A
  similar smoothness has been found for the transition to
  intermittency in the Pomeau-Manneville map \cite{KKCSG06}.}  In the
following we will compare these numerical results for the position
correlation decay with exact and approximate analytical results by
confirming our findings.

\subsection{Exact analytical results for small times}\label{subsec-k123}

\begin{figure}[H] 
	\centering
	\subfloat[$k = 1$]{
	\includegraphics[width = 0.27\linewidth]{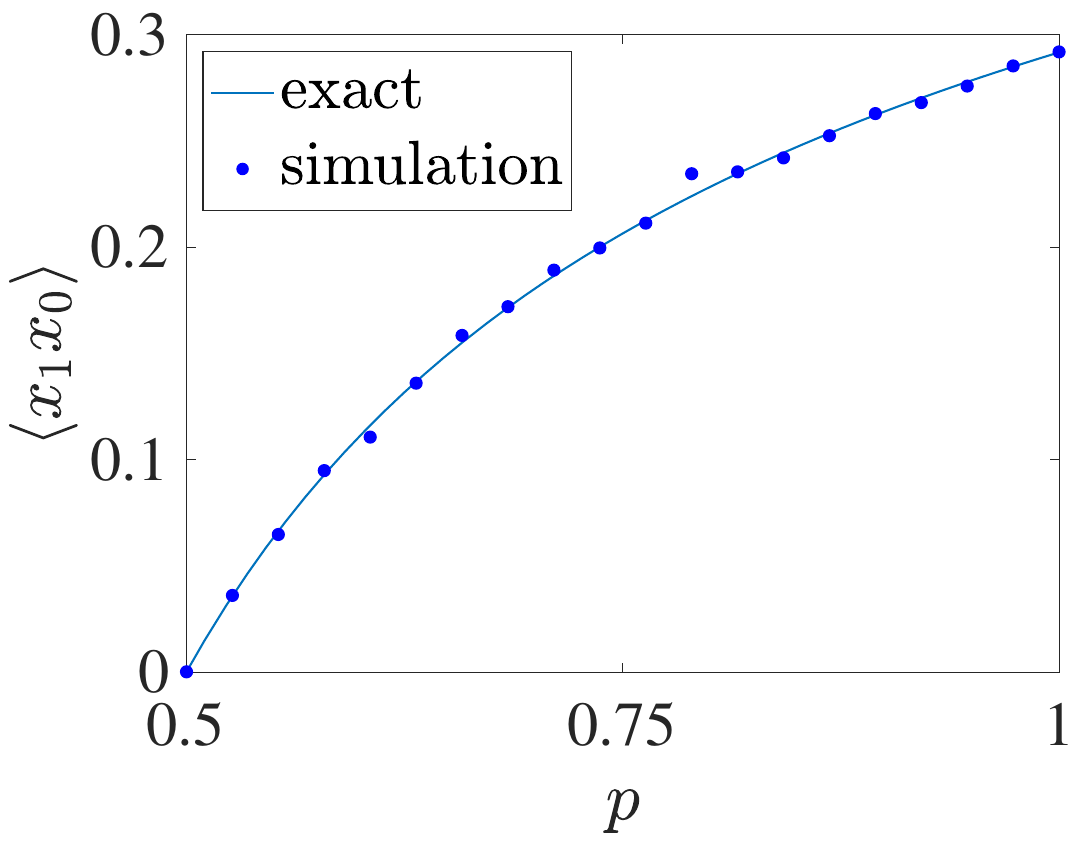}
	} 
	\subfloat[$k = 2$]{
	\includegraphics[width = 0.27\linewidth]{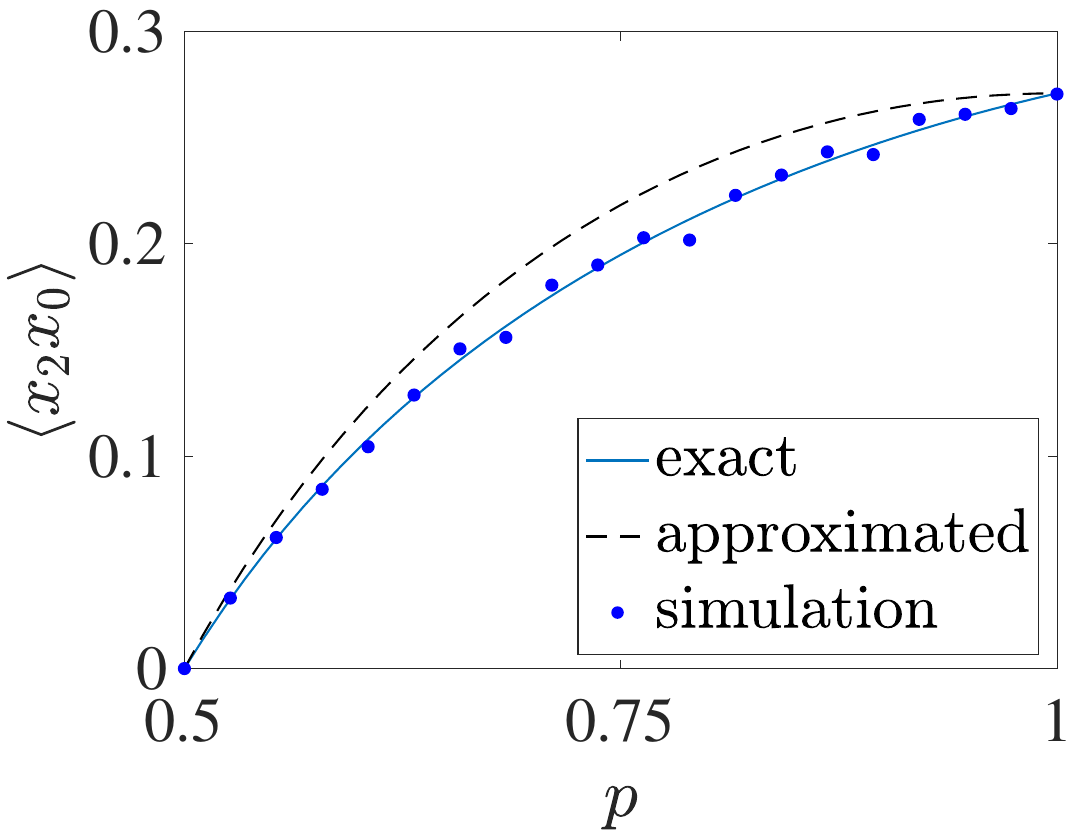}
	} 
	\subfloat[$k = 3$]{
	\includegraphics[width = 0.27\linewidth]{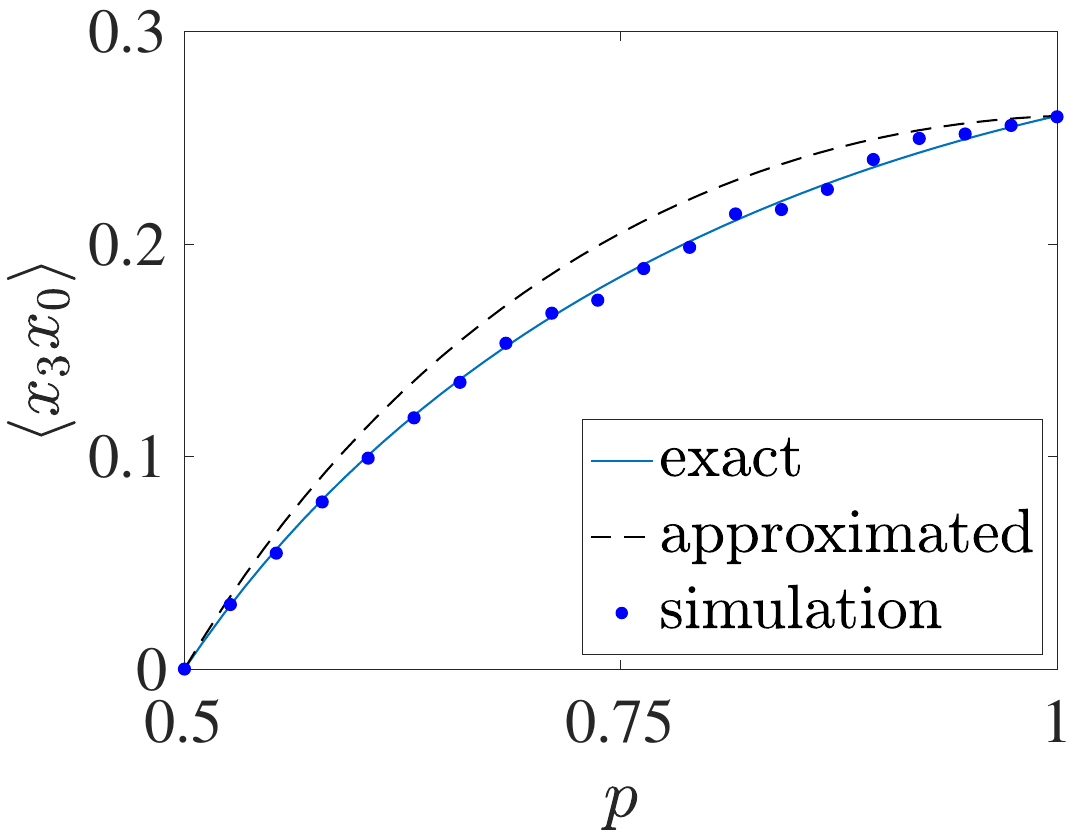}
	} 

\caption{Position autocorrelation function $\langle x_k x_0\rangle$
  Eq.~\eqref{eq:acf} for small times $k = 1, 2, 3$. In each plot, the
  solid curve represents the exact analytic result
  Eqns.~\eqref{x1x0-exact}, \eqref{x2x0-exact} and
  \eqref{x3x0-exact}. The scattered points depict the simulation data
  for which $10^5$ initial points have been sampled directly from the
  invariant density formula, with a truncation of $20$ subintervals on
  the unit interval, where $20$ equidistant values of $p \in [0.5001,
    0.9999]$ have been used. The two dashed curves show results from
  our analytical approximation Eqns.~\eqref{x2x0-semiM},
  \eqref{x3x0-semiM}, }

\label{k123plots}
\end{figure}

We now present exact analytical results for the autocorrelation decay
at the first three time steps. For $k = 1$, the first term of the
correlation function Eq.~\eqref{eq:acf} can be calculated to
\begin{equation}
\begin{split}
\langle x_1 x_0\rangle &= p \int_0^1 \rho_p(x_0) T_1(x_0)x_0 dx_0 + (1-p)\int_0^1 \rho_p(x_0) T_2(x_0)x_0 dx_0 \nonumber \\
&= 2p\langle x^2\rangle - a_0 p \int_{\frac{1}{2}}^1 x dx + \frac{1-p}{2}\langle x^2 \rangle \nonumber \\
&= \frac{(2p-1)(3p+25)}{24(5p-1)}\:,
\label{x1x0-exact}
\end{split}
\end{equation}
where we have used the second moment $\langle x^2 \rangle = \int
\rho_p(x) x^2 dx = \frac{4(2p-1)}{3(5p-1)}$
(App.~\ref{appendix-2nd-moment}). Deriving correlations for larger
times is more involved, since the maps $T_1$ and $T_2$ do not commute,
\begin{equation}
(T_2 T_1 - T_1T_2)(x) = \frac{1}{2}(2x \mod 1) - \left[ 2\left( \frac{x}{2}\right) \mod 1\right] = \begin{cases}
0, &\text{ if } x \in \left[0, \frac{1}{2}\right)\\
-\frac{1}{2}, &\text{ if } x \in \left[\frac{1}{2}, 1\right]\:.
\end{cases}
\label{commutator}
\end{equation}
Thus, here we only provide the next two examples of exact values for
$k=2,3$. By writing down all possibilities of the second iterate $x_2$
we obtain (App.~\ref{appendix-x2x0-x3x0-exact})
\begin{equation}
\langle x_2 x_0 \rangle = \int_0^1 \rho(x_0) x_2 x_0 dx_0
= (2p-1) \left[ \frac{(3p+1)^2}{3(5p-1)} - \frac{11p+6}{16}\right]. 
\label{x2x0-exact}
\end{equation}
It can be easily checked that at $p = 1$, $\langle x_2 x_0 \rangle_{p
  = 1} = 13/48$, which is consistent with the result for the
{doubling map} \cite{Yan21}. When $k = 3$ we have
(App.~\ref{appendix-x2x0-x3x0-exact})
\begin{equation}
\langle x_3 x_0 \rangle = \int_0^1 \rho(x_0) x_3 x_0 dx_0
= (2p-1)\left[ \frac{(3p+1)^3}{6(5p-1)} - \frac{28p^2 + 40p + 9}{32}\right]. 
\label{x3x0-exact}
\end{equation}
One can check again that this formula reproduces the result for the
{doubling map} at $p = 1$ \cite{Yan21}. In Fig.~\ref{k123plots} we
compare these three exact results with simulations. We see that the
agreement is excellent, confirming the quality of our simulations.

\subsection{Analytical approximations for larger times}

As explained in Sec.~\ref{subsec-k123}, the two maps defining the
Pelikan map do not commute, see Eq.~\eqref{commutator}, hence
the system is non-Markovian. However, the invariant density is highly
concentrated on the subinterval $x \in [0, 1/2)$
  when $p \to 1/2^+$, where the commutator
  Eq.~\eqref{commutator} is zero. In turn, for $p\to1$ the contraction
  $T_2$ is randomly chosen less frequently, thus generating fewer
  combinations of $T_i T_j$ ($i \neq j$). For these two limiting
  regimes of $p$ it is thus reasonable to approximate the original
  random system as if $T_1$ and $T_2$ were commutative, which is
  expected to yield a good qualitative approximation. In these
  situations we employ that iterates of the approximated map have a
  simple binomial structure. This enables us to derive analytically
  the autocorrelation function for a general time $k \in \mathbb{N}$.
  It should be noted, however, that our approach is only partially
  Markovian, in the sense that the original (non-commutative) dynamics
  is replaced by the Markovian dynamics. Yet, we are still assuming
  the same invariant density (i.e., the one for the actual
  non-commutative system).

First consider the second iteration of the approximated commutative
system. It can take three possible values, namely
\begin{equation}
x_2 =
\begin{cases}
T_1 (T_1 (x_0)) = 4x_0 \mod 1, &\quad \text{prob. } p^2\\
T_i (T_j (x_0)) \simeq x_0, \quad i \neq j, &\quad \text{prob. } 2p(1-p)\\
T_2 (T_2(x_0)) = \frac{1}{4}x_0, &\quad \text{prob. } (1-p)^2, 
\end{cases}
\label{x2-semiM}
\end{equation}
and thus (App.~\ref{appendix-appro-corre23})
\begin{equation}
\begin{split}
\langle x_2 x_0 \rangle
& \simeq  p^2\int_0^1 \rho_p (x_0)(4x_0 \mod 1)x_0 dx_0 + 2p(1-p)\int_0^1 \rho_p(x_0) x_0^2 dx_0 + (1-p)^2 \int_0^1 \rho_p(x_0)\frac{1}{4}x_0^2 dx_0\\
&= (2p-1)\left[ \frac{(3p+1)^2}{3(5p-1)} - \frac{14p+3}{16}\right].
\end{split}
\label{x2x0-semiM}
\end{equation}
Similarly, we obtain 
\begin{equation}
\langle x_3 x_0 \rangle \simeq  (2p-1)\left[ \frac{(3p+1)^3}{6(5p-1)} - \frac{27p^2 + 47p + 3}{32}\right]. 
\label{x3x0-semiM}
\end{equation}

Figure~\ref{k123plots} compares these approximate results with both
the exact and the numerical values for small times $k = 1, 2, 3$. We
see that, the approximation qualitatively reproduces the functional
form of the autocorrelation function under variation of $p$. As
expected, the quantitative deviations are largest for intermediate
values of $p$ while the approximation works reasonably well for the
limiting cases of $p\to1/2$ and $p\to1$.

For general times $k \in \mathbb{N}$, $\langle x_k x_0 \rangle$ can be
obtained by writing out all $k+1$ possible values of the $k$th
iteration with corresponding binomial probabilities ${k \choose j}p^j
(1-p)^{k-j}$, $j = 0, 1, ..., k$. A lengthy calculation yields
\cite{Yan21}
\begin{equation}
\begin{split}
&\langle x_kx_0\rangle (p) \\
 \simeq & \frac{4(2p-1)}{3(5p-1)}\left(\frac{3p+1}{2}\right)^k \\
  &- \frac{2p-1}{2}\left\{ \frac{8}{3(5p-1)}\left[ \left(\frac{3p+1}{2}\right)^k - (2p)^k \left(\frac{1-p}{4p}\right)^{K+1}{k \choose K+1}\cdot {}_2F_1 \left(a_1, a_2; b; -\frac{1-p}{4p} \right)\right] \right. \\
&\left. - \frac{1}{3p-1}\left[ 1 - p^k\left(\frac{1-p}{p}\right)^{K+1}{k \choose K+1}\cdot {}_2F_1 \left(a_1, a_2; b; -\frac{1-p}{p} \right)\right] \right. \\
&\left. -\frac{1}{6(2p-1)}\left[ \left(\frac{4-3p}{2}\right)^k - \left(\frac{p}{2}\right)^k\left(\frac{4(1-p)}{p}\right)^{K+1}{k \choose K+1}\cdot {}_2F_1 \left(a_1, a_2; b; -\frac{4(1-p)}{p} \right)\right] \right. \\
&\left. - \frac{(7p-3)(p-1)}{2(2p-1)(3p-1)(5p-1)}\left[\left(\frac{3p+1}{2}\right)^k - \left(\frac{1-p}{2}\right)^k\left(\frac{4p}{1-p}\right)^{K+1}{k \choose K+1}\cdot {}_2F_1 \left(a_1, a_2; b; -\frac{4p}{1-p} \right)\right]\right\},
\end{split}
\label{2pt-corre-v2}
\end{equation}
where ${}_2F_1(a_1, a_2; b; x)$ is the Gauss hypergeometric function,
$a_1 = 1$, $a_2 = -k+K+1$, $b = K+2$ and, $K = \frac{k}{2} - 1$ for
even $k$ and $K = \frac{k-1}{2}$ for odd $k$. The difference for the
results of odd and even $k$ is due to the fact that, for even $k$,
there is a possibility that the approximated map becomes the identity,
where we have $x_k = x_0$. But for odd $k$ this is impossible, as each
odd iteration either expands or contracts the system.

\begin{figure}[H]
\centering
\subfloat[$p = 0.9999$]{
	\includegraphics[width = 0.36\linewidth]{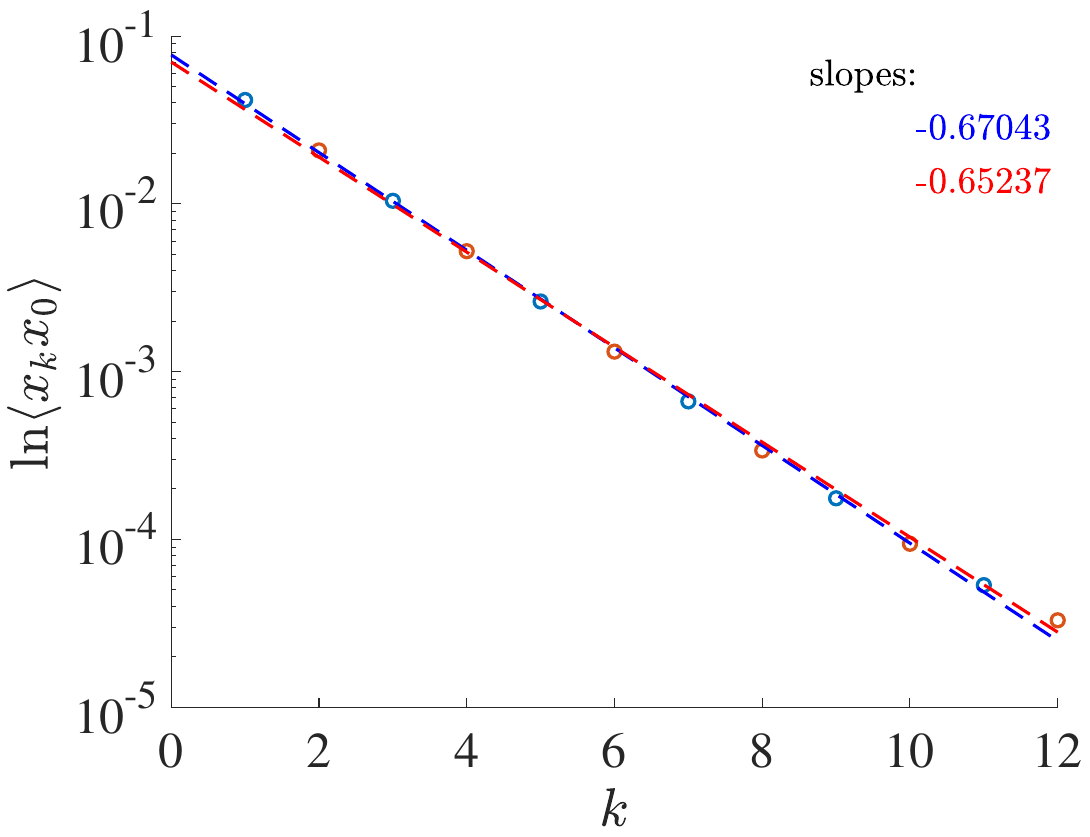}
	\label{analytic-semilog}
}
\subfloat[$p = 0.5001$]{
	\includegraphics[width = 0.36\linewidth]{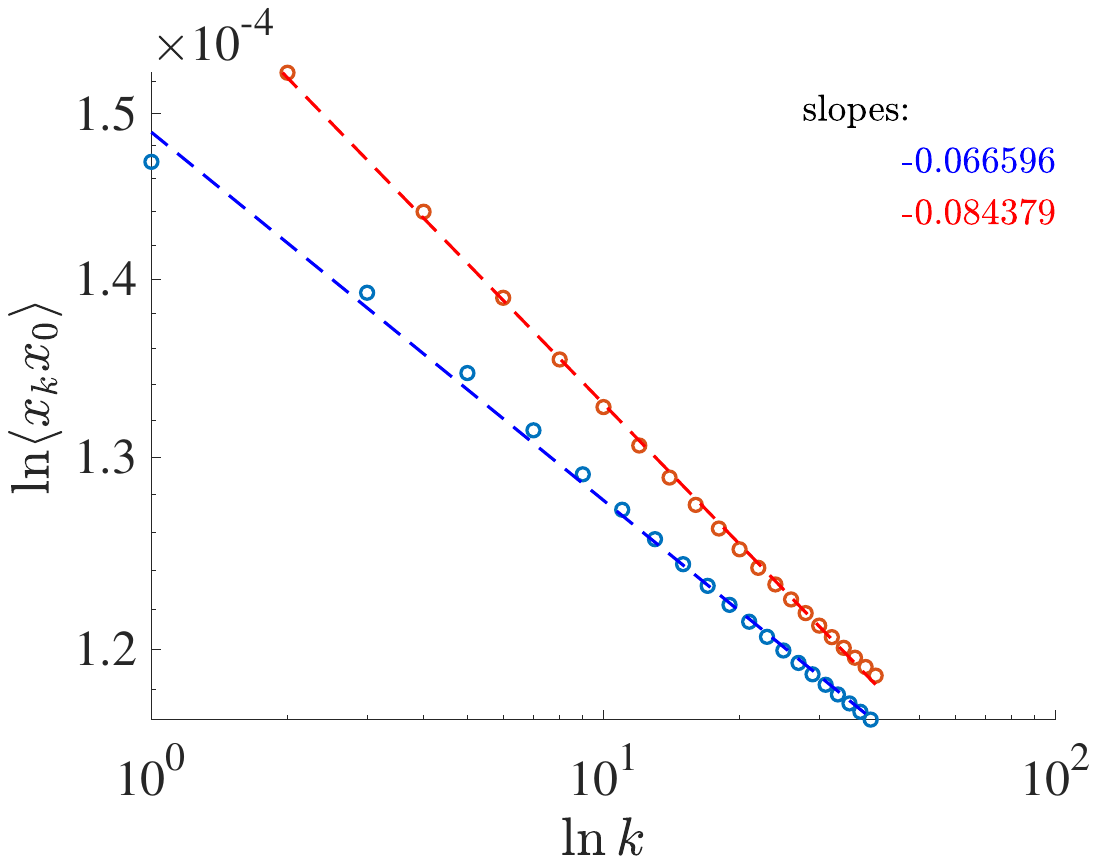}
	\label{analytic-loglog}
}
\caption{Position autocorrelation functions $\langle x_k x_0 \rangle$
  Eq.~\eqref{2pt-corre-v2} for (a) $p = 0.9999, k = 1, 2, ..., 12$
  (semi-logarithmic scale), (b) $p = 0.5001, k = 1, 2, ..., 40$
  (double-logarithmic scale). Odd times $k$ are depicted in blue, even
  $k$ in red. Dashed lines are linear fits with the slopes indicated
  on the top-right corner.}
\label{corr-analytic}
\end{figure}

We plot the results for the autocorrelation decay obtained from this
formula in Fig.~\ref{corr-analytic}. We see in
Fig.~\ref{corr-analytic}(a) that for $p \to 1^-$ our analytical
approximation correctly reproduces an exponential decay for odd and
even $k$. This is consistent with the fact that for the {doubling
  map} (i.e., $p = 1$) we have precisely $\langle \tilde{x}_k
\tilde{x}_0 \rangle = (1/12)\left( 1/2\right)^k$
\cite{MSO81,yan2020distinguished,Yan21}. In semi-logarithmic
representation this yields a slope of $\ln(1/2) \approx -0.69$, which
matches well to the results of our analytical
approximation. Figure~\ref{corr-analytic}(b) confirms analytically the
power-law correlation decay observed in our simulation results of
Fig.~\ref{nCFplot}(b).

We remark that at each iteration step in Eq.~\eqref{2pt-corre-v2} a
small deviation is made by this approximated approach compared to the
exact Pelikan map. In the long run (i.e., $k \to \infty$) these errors
will accumulate so that the exact results for the correlation decay of
the Pelikan map will substantially deviate from this approximation.
Our approximation is thus only valid for smaller times $k$, and in the
two limits $p \to 1/2^+$, $p \to 1^-$. When $p$ is between
these two limits, the non-commutativity of the Pelikan map plays an
important role in the dynamics. How to better capture this property
analytically is an interesting open question.

\section{Conclusion and outlook}
\label{sec-outlook}

In this paper we studied a transition scenario that is generated by a
class of simple random dynamical systems, i.e., piecewise linear
one-dimensional maps, where we focused on the Pelikan map. This
particular model stands out as one of the rare examples that can be
understood analytically for the entire range of parameter values. The
transition exhibited by it also stands out, because it is of
particular interest to learn how a typical transition scenario looks
like in a system where a parameter is changed that controls the
relative likelihood by which an expanding dynamics is chosen as
compared to a contracting one \cite{SaKl19}.

To characterise this transition, we first derived analytically the
coarse-grained invariant density of the Pelikan map and found that it
undergoes multiple transitions under parameter variation, such as
changes in convexity, boundedness and normalisability. All these
critical behaviours can be extended to a broader family of random maps
that includes the Pelikan map as a special case. Secondly, we provided
numerical evidence for weak ergodicity breaking at the critical
parameter value where the normalisation of the invariant density
changes from finite to infinite. This was demonstrated by an
inequality between time and ensemble average of the position as an
observable. Thirdly, we studied the position autocorrelation function
both numerically and analytically. Our findings unveiled a transition
from exponential to power-law decay under parameter variation,
indicating the emergence of anomalous dynamics at the transition
point. Overall, our study thus provides a comprehensive examination of
the complete dynamical transition in the Pelikan map under variation
of the sampling probability as a parameter. Characteristic properties
such as critical density changes, weak ergodicity breaking and
anomalous correlation decay are expected to be observed in other, more
complex dynamical systems as well.

There are many interesting open questions. Perhaps the most important
one is the derivation of an exact analytical expression for the
position autocorrelation function of the Pelikan map at the transition
point for all times, respectively a mathematical proof of its decay in
the limit of long times. To answer this question it would be
interesting to explore more sophisticated methods expressing the
correlation function in terms of eigenvalues and eigenfunctions of the
Frobenius-Perron operator of the random map, by then solving the
associated eigenvalue problem, as has been accomplished for simpler
deterministic maps \cite{MSO81}. Although many features of the Pelikan
map can be analytically understood more easily, this particular
problem turned out to be very difficult because of the non-commuting
properties that we highlighted in Sec.~\ref{subsec-k123}. Notably,
many phenomena in quantum systems involve non-commuting operators, so
from this point of view it would be interesting to make further
progress in this direction. This would be particularly significant in
the realm of other non-commutative dynamical systems as well, being
quantum mechanical or just classical, as long as they include a random
sampling of different dynamics where the order of the choice is
relevant.

It would also be interesting to further employ this class of maps for
defining associated diffusive dynamical systems. This has already been
done for a particular setting in Ref.~\cite{SaKl19} by delivering an
interesting transition from normal to anomalous
(sub)diffusion. However, it would be important to study yet other
types of diffusive random dynamical systems. Especially, obtaining one
that generates superdiffusion is currently an open question.  In
Ref.~\cite{beck1996dynamical} a deterministic Langevin dynamics has
been proposed, where the iterates of a simple one-dimensional map are
used as the fast driving force for a slowly changing dissipative
dynamics. In other words, the Gaussian white noise of an ordinary
Langevin equation has been replaced by a more complex dynamical
systems noise source, for which also the Pelikan map could be
used. This spatially extended dynamical systems would be diffusive and
would allow us to investigate the influence of the transition
controlled by the sampling probability of the Pelikan map onto a
corresponding diffusion processes (defined, in the simplest case, by
the variable $y_n$ as given in Eq.~\eqref{eq-yn}).  It has been
suggested that strongly chaotic behaviour in dynamical systems
generates normal diffusion, whereas non-stationary intermittency (or
weakly chaotic dynamics) yields anomalous diffusion
\cite{Kla13,SaKl19}. A diffusive dynamical system driven by the
Pelikan map would thus provide an interesting test ground to further
investigate these types of questions. Our observations reported in
this paper indicate that in the vicinity of the critical point,
represented by the parameter value where the Lyapunov exponent is
zero, the decay of the position autocorrelation function behaves
anomalously. Correspondingly, an associated diffusive random dynamical
system should generically display a transition to anomalous diffusion.

So far, our investigations have focused solely on one-dimensional
random maps. However, it is worth noting that higher-dimensional
random dynamical systems of the same type can also be defined, such
as, e.g., iterated function systems \cite{Barn93}, which may exhibit
significantly more complex transitions. Exploring higher-dimensional
random dynamical systems could potentially provide deeper insights
into more complex real-world systems. For such an endeavour the
Pelikan map promises to serve as an important guideline, in the form
of a low-dimensional example where many generic phenomena can already
be understood in detail.\\

{\bf Acknowledgement:} The authors thank the London Mathematical
Laboratory for financial support and the initialisation of this
project in the form of two Summer Schools at the ICTP Trieste in 2019
and 2021. Here YS, SR and RK worked together with JY and MM during
one-month Summer projects, leading to the present results. RK and YS
thank Prof.~A.J.Homburg for discussions, which started at a workshop
in 2019 where we presented our work Ref.~\cite{SaKl19}, as well as for
making them aware of Refs.~\cite{HoKa22,Hom23}. RK also thanks
Dr.~W.Just, Dr.~M.Rasmussen and Prof.~J.S.W.Lamb for hints on relevant
literature and many helpful remarks.

\newpage
\appendix
\renewcommand{\thesection}{\Alph{section}}
\section{Derivation of the invariant density curve formula}
\label{appendix-inv-den-curve}
For $p \in \left(\frac{1}{2}, 1\right]$, the invariant density is
normalisable, i.e., we have $\sum_{i = 0}^{\infty}r_i = 1$,
\begin{equation*}
\begin{split}
1 &= \lim_{n \rightarrow \infty}(r_0 + r_1 + ... + r_n) \\
&= \lim_{n \rightarrow \infty}\left( \frac{1}{2}a_0 + \frac{1}{2^2}R_1a_0 + ... + \frac{1}{2^{n+1}}R_n \cdot R_{n-1}\cdot ... \cdot R_1a_0\right) \\
&= \frac{a_0}{2} \lim_{n \rightarrow \infty} \left( 1 + \frac{1}{2}R_1 + \frac{1}{2^2}R_2R_1 + ... + \frac{1}{2^n}R_n \cdot ... \cdot R_1\right) \\
&= \frac{a_0}{-1 + \frac{2(1-p)}{p}}\lim_{n\rightarrow \infty} \left( -\frac{1}{2} + \frac{1-p}{p} - \frac{1}{4} \sum_{i = 0}^{n-1}\left(\frac{1}{2}\right)^i + \left( \frac{1-p}{p}\right)^2\sum_{i = 0}^{n-1}\left(\frac{1-p}{p}\right)^{n-1}\right) \\
&= \frac{a_0}{-1 + \frac{2(1-p)}{p}}\left( -\frac{1}{2} + \frac{1-p}{p} - \frac{1}{4}\cdot \frac{1}{1 - \frac{1}{2}} + \left(\frac{1-p}{p}\right)^2 \cdot \frac{1}{1 - \frac{1-p}{p}}\right) \\
&= \frac{a_0 p}{2p-1},
\end{split}
\end{equation*}
where from the second line onwards we have denoted
\begin{equation*}
R_n = R_n(p) := \frac{a_n}{a_{n-1}} = 2\frac{r_n}{r_{n-1}} = \frac{-1 + \left(\frac{2(1-p)}{p}\right)^{n+1}}{-1 + \left(\frac{2(1-p)}{p}\right)^n}.
\end{equation*}
Therefore,
\begin{equation*}
a_0 = \frac{2p - 1}{p}, \quad p \in \left(\frac{1}{2}, 1\right]. 
\end{equation*}
By the recurrence relation of $\{a_n\}$ we have an explicit form for the amplitude 
\begin{equation}
a_n = \frac{-1 + \left(\frac{2(1-p)}{p}\right)^{n+1}}{-1 + \frac{2(1-p)}{p}}a_0 = \frac{2p-1}{2-3p}\left[-1 + \left(\frac{2(1-p)}{p}\right)^{n+1}\right].
\label{an1}
\end{equation}
The invariant density curve is given by the mid-point interpolation
$\left( \frac{3}{2^{n+2}}, a_n\right)$, and with Eq.~\eqref{an1},
\begin{equation*}
\begin{split}
x &= \frac{3}{2^{n+2}} \quad \Rightarrow n+1 = \log_2 \frac{3}{2x} = \frac{\ln 3}{\ln 2} - 1 - \frac{\ln x}{\ln 2}, \\
\tilde{\rho}_p(x) &= \frac{2p-1}{3p-2}\left[ 1 - \left( \frac{2(1-p)}{p}\right)^{n+1}\right] \\
&= \frac{2p-1}{3p-2}\left[ 1 - \left( \frac{2(1-p)}{p}\right)^{\frac{\ln 3}{\ln 2} - 1 - \frac{\ln x}{\ln 2}}\right] \\
&= \frac{2p-1}{3p-2}\left[ 1 - \left( \frac{2(1-p)}{p}\right)^{\frac{\ln 3}{\ln 2} - 1} \left( \frac{p}{2(1-p)}\right)^{\frac{\ln x}{\ln 2}}\right] \\
&= \frac{2p-1}{3p-2}\left[ 1 - \left( \frac{2(1-p)}{p}\right)^{\frac{\ln 3}{\ln 2} - 1} e^{\frac{\ln x}{\ln 2}\ln \frac{1}{2}} \left( \frac{p}{1-p}\right)^{\frac{\ln x}{\ln 2}} \right] \\
&= \frac{2p-1}{3p-2}\left[ 1 - \left( \frac{2(1-p)}{p}\right)^{\frac{\ln 3}{\ln 2} - 1} x^{-1} e^{\ln x \frac{1}{\ln 2}\ln \frac{p}{1-p}} \right] \\
&= \frac{2p-1}{3p-2}\left[ 1 - \left( \frac{2(1-p)}{p}\right)^{\frac{\ln 3}{\ln 2} - 1} x^{-1 + \frac{1}{\ln 2}\ln \frac{p}{1-p}} \right] \\
&=: A(1-Bx^{-1+C}), \\
& \text{ where } A(p) := \frac{2p-1}{3p-2}, B(p) := \left( \frac{2(1-p)}{p}\right)^{\frac{\ln 3}{\ln 2} - 1}, C(p) := \frac{1}{\ln 2}\ln \frac{p}{1-p}.
\end{split}
\end{equation*}

\section{Derivation of the convexity change for a general case}
\label{appendix-convexity}

The method follows directly from Ref.~\cite{Pel84} except that the
expansion rate is any integer $s\geqslant 2$.  Analogous to
Eq.\eqref{relation-r-a}, for $s_1 = s = 1/s_2$ we have
\begin{equation}
\frac{r_{n+1}}{r_n} = \frac{-1 + \left( \frac{s(1-p)}{p}\right)^{n+2}}{-s + s\left( \frac{s(1-p)}{p}\right)^{n+1}}. 
\end{equation}
The normalisation condition 
\begin{equation}
1 = \sum_{j=0}^{\infty}r_j = \lim_{n \to \infty} (r_0 + r_1 + ... + r_n) = a_0 \frac{p}{(s-1)(2p-1)}
\end{equation}
gives $a_0 = \frac{(s-1)(2p-1)}{p}$ and therefore 
\begin{equation}
a_n = \frac{-1+\left(\frac{s(1-p)}{p}\right)^{n+1}}{-1 + \frac{s(1-p)}{p}} a_0 = \frac{(s-1)(2p-1)}{s - (s+1)p}\left[ -1 + \left( \frac{s(1-p)}{p}\right)^{n+1}\right].
\end{equation}
By mid-point interpolation $\left( \frac{s+1}{2s^{n+1}}, a_n\right)$ (cf. Sec.~\ref{appendix-inv-den-curve}) we get 
\begin{equation}
\tilde{\rho}_p(x) = A_s (1 - B_s x^{-1+C_s}), 
\end{equation}
where 
\begin{equation}
A_s(p) = \frac{(s-1)(2p-1)}{(s+1)p - s}, \quad 
B_s(p) = \left( \frac{s(1-p)}{p}\right)^{\frac{\ln(s+1)}{\ln s} - \frac{\ln 2}{\ln s}}, \quad 
C_s(p) = \frac{1}{\ln s} \ln \frac{p}{1-p}.
\end{equation}
When $s = 2$ we recover the Pelikan case.  The convexity changes when
$C_s = 2$, which gives the critical value of $p$: $p_c = \frac{s^2}{1
  + s^2}$.

\section{Derivation of the moments of the Pelikan map}
\label{appendix-2nd-moment}
The $m$th moment is given by the expectation of $x^m$ with respect to the invariant density $\rho_p(x)$ of the Pelikan map \cite{Yan21},
\begin{equation*}
\begin{split}
\langle x^m \rangle &= \int_0^1\rho_p(x)x^m dx \\
&= \sum_{j = 0}^{\infty}a_j \int_{\frac{1}{2^{j+1}}}^{\frac{1}{2^j}}x^m dx \\
&= \frac{1}{m+1}\sum_{j = 0}^{\infty}a_j \left( \frac{1}{2^{j(m+1)}} - \frac{1}{2^{(j+1)(m+1)}}\right)\\
&= \frac{1}{m+1}\cdot\left( 1 - \frac{1}{2^{m+1}}\right) \sum_{j = 0}^{\infty}a_j \left(\frac{1}{2^{m+1}}\right)^j\\
&= \frac{1}{m+1}\cdot \left( 1 - \frac{1}{2^{m+1}}\right)\cdot \frac{2p-1}{3p-2} \sum_{j = 0}^{\infty} \left[ 1 - \left(\frac{2(1-p)}{p}\right)^{j+1}\right] \left(\frac{1}{2^{m+1}}\right)^j\\
&= \frac{1}{m+1}\cdot \left( 1 - \frac{1}{2^{m+1}}\right)\cdot \frac{2p-1}{3p-2} \left[\sum_{j = 0}^{\infty}\left(\frac{1}{2^{m+1}}\right)^j - \frac{2(1-p)}{p}\sum_{j = 0}^{\infty}\left(\frac{1-p}{2^m p}\right)^j\right]\\
&= \frac{1}{m+1}\cdot \frac{2^{m+1} - 1}{2^{m+1}}\cdot \frac{2p-1}{3p-2} \left[\frac{2^{m+1}}{2^{m+1} - 1} - \frac{2^{m+1}(1-p)}{2^m p - 1 + p}\right]\\
&= \frac{1}{m+1}\cdot \frac{2p - 1}{3p - 2}\cdot \frac{2^m (3p - 2)}{2^m p - 1 + p}\\
&= \frac{2^m}{m+1}\cdot \frac{2p - 1}{(2^m + 1)p - 1}.
\end{split}
\end{equation*}
In particular, when $m = 1$, the average of $x$ reads 
\begin{equation*}
\langle x \rangle = \frac{2p-1}{3p-1}, 
\end{equation*}
and when $m = 2$, 
\begin{equation*}
\langle x^2 \rangle 
= \frac{4(2p-1)}{3(5p-1)}.
\end{equation*}

\section{Derivation of the exact auto-correlation functions of time differences $k = 2$ and $3$}
\label{appendix-x2x0-x3x0-exact}
The second iterate of $x$ can take different values with their corresponding probabilities: 
\begin{equation}
x_2 = \begin{cases}
T_1(T_1(x_0)) = 4x_0 \text{ mod }1 = \begin{cases}
4x_0, &x_0 \in \left[ 0, \frac{1}{4}\right)\\
4x_0 - 1, &x_0 \in \left[ \frac{1}{4}, \frac{1}{2}\right)\\
4x_0 - 2, &x_0 \in \left[ \frac{1}{2}, \frac{3}{4}\right)\\
4x_0 - 3, &x_0 \in \left[ \frac{3}{4}, 1\right]
\end{cases}
, &\text{ prob. } p^2\\
T_2(T_1(x_0)) = \frac{1}{2}(2x_0 \text{ mod }1) = 
\begin{cases}
x_0, \quad &x_0 \in \left[0, \frac{1}{2}\right)\\
x_0 - \frac{1}{2}, \quad &x_0 \in \left[ \frac{1}{2}, 1\right]
\end{cases}, &\text{ prob. } (1-p)p\\
T_1(T_2(x_0)) = x_0, &\text{ prob. } p(1-p)\\
T_2(T_2(x_0)) = \frac{1}{4}x_0, &\text{ prob. } (1-p)^2
\end{cases}
\label{x2-exact}
\end{equation}
Therefore,
\begin{equation*}
\begin{split}
\langle x_2 x_0 \rangle =& \int_0^1 \rho(x_0) x_2 x_0 dx_0\\
=& p^2 \int_0^1 \rho(x)(4x \text{ mod }1)x dx + p(1-p)\left[ \int_0^{\frac{1}{2}}\rho(x) x^2 dx + \int_{\frac{1}{2}}^1 \rho(x) (x - \frac{1}{2})x dx\right]\\
&+ p(1-p) \int_0^1 \rho(x) x^2 dx + (1-p)^2\int_0^1 \rho(x)\frac{1}{4}x^2dx\\
=& p^2 \left[ \int_0^{\frac{1}{4}} \rho(x) 4x^2dx + \int_{\frac{1}{4}}^{\frac{2}{4}} \rho(x) (4x-1)x dx + \int_{\frac{2}{4}}^{\frac{3}{4}}\rho(x) (4x-2)x dx + \int_{\frac{3}{4}}^1 \rho(x) (4x-3)xdx \right] + \\
&+ p(1-p)\left[ \int_0^{\frac{1}{2}} \rho(x)x^2dx + \int_{\frac{1}{2}}^1 \rho(x)(x^2 - \frac{1}{2}x)dx\right] + p(1-p) \langle x^2 \rangle + \frac{(1-p)^2}{4}\langle x^2 \rangle \\
=& 4p^2 \langle x^2\rangle - p^2 \left[ \int_{\frac{1}{4}}^{\frac{2}{4}}\rho(x)xdx + 2\int_{\frac{2}{4}}^{\frac{3}{4}} \rho(x) xdx + 3\int_{\frac{3}{4}}^1 \rho(x)xdx \right] + p(1-p)\left[ \langle x^2 \rangle - \frac{1}{2}\int_{\frac{1}{2}}^1 \rho(x) xdx\right] \\
&+ p(1-p)\langle x^2 \rangle + \frac{(1-p)^2}{4}\langle x^2 \rangle\\
=& \left( 4p^2 + 2p(1-p) + \frac{(1-p)^2}{4}\right) \langle x^2\rangle - p^2 \left[ a_1 \left.\frac{x^2}{2}\right\rvert_{\frac{1}{4}}^{\frac{2}{4}} + 2a_0 \left.\frac{x^2}{2}\right\rvert_{\frac{2}{4}}^{\frac{3}{4}} + 3a_0 \left.\frac{x^2}{2}\right\rvert_{\frac{3}{4}}^1\right] - \frac{p(1-p)}{2}a_0 \left. \frac{x^2}{2}\right\vert_{\frac{1}{2}}^1\\
=& \frac{(3p+1)^2}{4}\langle x^2 \rangle - p^2 \left[ \frac{(2p-1)(2-p)}{2p^2}\frac{4 - 1}{16} + 2\frac{2p-1}{2p}\frac{9 - 4}{16} + 3\frac{2p-1}{2p}\frac{16 - 9}{16}\right] - \frac{p(1-p)}{4}\frac{2p-1}{p}\frac{4 - 1}{4}\\
=& \frac{(3p+1)^2}{4} \frac{4(2p-1)}{3(5p-1)} - \frac{(2p-1)(14p+3)}{16} - \frac{3(2p-1)(1-p)}{16} \\ 
=& (2p-1) \left[ \frac{(3p+1)^2}{3(5p-1)} - \frac{11p+6}{16}\right]. 
\end{split}
\end{equation*}

The third iteration of the Pelikan map reads
\begin{equation*}
x_3 = \begin{cases}
\begin{cases}
T_1 T_1 T_1 (x_0) = 8x_0 \text{ mod }1 = \textit{ [8 cases]}, \quad & \text{prob. $p^3$}\\
T_2 T_1 T_1 (x_0) = \frac{1}{2}(4x_0 \text{ mod }1) = 
\begin{cases} 
\frac{1}{2}(4x_0) = 2x_0, \quad &x_0 \in \left[ 0, \frac{1}{4}\right)\\
\frac{1}{2}(4x_0 - 1) = 2x_0 - \frac{1}{2}, \quad &x_0 \in \left[ \frac{1}{4}, \frac{1}{2}\right)\\
\frac{1}{2}(4x_0 - 2) = 2x_0 - 1, \quad &x_0 \in \left[ \frac{1}{2}, \frac{3}{4}\right)\\
\frac{1}{2}(4x_0 - 3) = 2x_0 - \frac{3}{2}, \quad &x_0 \in \left[ \frac{3}{4}, 1\right]
\end{cases}, \quad &\text{ prob. $(1-p)p^2$}\\
\end{cases}\\
\begin{cases}
T_1 T_2 T_1 (x_0) = 2x_0 \text{ mod }1 = \textit{ [2 cases]}, \quad &\text{prob. $p(1-p)p$}\\
T_2 T_2 T_1 (x_0) = \frac{1}{4}(2x_0 \text{ mod }1) = 
\begin{cases}
\frac{1}{2}x_0, \quad &x_0 \in \left[0, \frac{1}{2}\right)\\
\frac{1}{2}x_0 - \frac{1}{4}, \quad &x_0 \in \left[ \frac{1}{2}, 1\right]
\end{cases}, \quad &\text{ prob. $(1 - p)^2 p$}
\end{cases}\\
\begin{cases}
T_1 T_1 T_2 (x_0) = 2x_0 \text{ mod }1 = \textit{ [2 cases]}, \quad &\text{prob. $p^2 (1-p)$}\\
T_2 T_1 T_2 (x_0) = \frac{1}{2}x_0, \quad &\text{prob. $(1-p)p(1-p)$}
\end{cases}\\
\begin{cases}
T_1 T_2 T_2 (x_0) = \frac{1}{2}x_0, \quad &\text{prob. $p(1-p)^2$}\\
T_2 T_2 T_2 (x_0) = \frac{1}{8}x_0, \quad &\text{prob. $(1-p)^3$}
\end{cases}
\end{cases}
\end{equation*}
which gives 
\begin{equation*}
\begin{split} 
\langle x_3 x_0 \rangle =& \int_0^1 \rho(x_0) x_3 x_0 dx_0 \\
=& p^3 \int_0^1 \rho(x) (8x \text{ mod }1)x dx + p^2(1-p)\left[ \int_0^{\frac{1}{4}}\rho(x)2x^2dx + \int_\frac{1}{4}^{\frac{1}{2}} \rho(x) (2x-\frac{1}{2})xdx + \int_{\frac{1}{2}}^{\frac{3}{4}}\rho(x) (2x-1)xdx \right.\\
& \left. + \int_{\frac{3}{4}}^1 \rho(x) (2x-\frac{3}{2})xdx \right] + 2p^2(1-p)\int_0^1 \rho(x) (2x \text{ mod }1)xdx + p(1-p)^2 \left[ \int_0^{\frac{1}{2}} \rho(x) \frac{1}{2}x^2dx + \int_{\frac{1}{2}}^1 \rho(x) (\frac{1}{2}x - \frac{1}{4})xdx\right] \\
& + 2p(1-p)^2 \int_0^1 \rho(x) \frac{1}{2}x^2 dx + (1-p)^3 \int_0^1 \rho(x) \frac{1}{8}x^2dx\\
=& p^3 \left[ 8\langle x^2\rangle - \int_{\frac{1}{8}}^{\frac{2}{8}} \rho(x)xdx - 2\int_{\frac{2}{8}}^{\frac{3}{8}} \rho(x)xdx - \int_{\frac{3}{8}}^{\frac{4}{8}} 3\rho(x)xdx - 4\int_{\frac{4}{8}}^{\frac{5}{8}} \rho(x)xdx - 5\int_{\frac{5}{8}}^{\frac{6}{8}} \rho(x)xdx - 6\int_{\frac{6}{8}}^{\frac{7}{8}} \rho(x)xdx \right. \\
&\left. - 7\int_{\frac{7}{8}}^1 \rho(x)xdx\right] + p^2(1-p)\left[ 2\langle x^2 \rangle - \frac{1}{2}\int_{\frac{1}{4}}^{\frac{1}{2}} \rho(x)xdx - \int_{\frac{1}{2}}^{\frac{3}{4}} \rho(x)xdx - \frac{3}{2}\int_{\frac{3}{4}}^1 \rho(x)dx \right] \\
&+ 2p^2(1-p) \left[ 2\langle x^2 \rangle - \int_{\frac{1}{2}}^1 \rho(x)xdx\right] + p(1-p)^2 \left[ \frac{1}{2}\langle x^2\rangle - \frac{1}{4}\int_{\frac{1}{2}}^1 \rho(x)xdx\right] + p(1-p)^2 \langle x^2\rangle + \frac{(1-p)^3}{8}\langle x^2 \rangle\\
=& \left( 8p^3 + 2p^2(1-p) + 4p^2(1-p) + \frac{p(1-p)^2}{2} + p(1-p)^2 + \frac{(1-p)^3}{8}\right) \langle x^2\rangle \\
&- p^3 \left[ \left. a_2\frac{x^2}{2}\right\vert_{\frac{1}{8}}^{\frac{2}{8}} + 2a_1\left. \frac{x^2}{2}\right\vert_{\frac{2}{8}}^{\frac{3}{8}} + 3a_1\left. \frac{x^2}{2}\right\vert_{\frac{3}{8}}^{\frac{4}{8}} + 4a_0\left. \frac{x^2}{2}\right\vert_{\frac{4}{8}}^{\frac{5}{8}} + 5a_0\left. \frac{x^2}{2}\right\vert_{\frac{5}{8}}^{\frac{6}{8}} + 6a_0\left. \frac{x^2}{2}\right\vert_{\frac{6}{8}}^{\frac{7}{8}} + 7a_0\left. \frac{x^2}{2}\right\vert_{\frac{7}{8}}^1 \right] \\
&- p^2(1-p)\left[ \frac{a_1}{2}\left. \frac{x^2}{2}\right\vert_{\frac{1}{4}}^{\frac{1}{2}} + a_0\left. \frac{x^2}{2}\right\vert_{\frac{1}{2}}^{\frac{3}{4}} + \frac{3a_0}{2}\left. \frac{x^2}{2}\right\rvert_{\frac{3}{4}}^1\right] - 2p^2(1-p) a_0\left. \frac{x^2}{2}\right\rvert_{\frac{1}{2}}^1 - \frac{p(1-p)^2}{4}a_0\left. \frac{x^2}{2}\right\rvert_{\frac{1}{2}}^1\\
=& \frac{(3p+1)^3}{8} \frac{4(2p-1)}{3(5p-1)} - \frac{p^3}{2}\left[ \frac{(2p-1)(3p^2 - 6p + 4)}{p^3}\frac{3}{64} + \frac{(2p-1)(2-p)}{p^2}\left( \frac{2\cdot 5}{64} + \frac{3\cdot 7}{64}\right) \right.\\
&\left. + \frac{(2p-1)}{p}\left( \frac{4\cdot 9}{64} + \frac{5\cdot 11}{64} + \frac{6\cdot 13}{64} + \frac{7\cdot 15}{64}\right)\right] - \frac{p^2(1-p)}{4}\left[ \frac{(2p-1)(2-p)}{p^2}\frac{3}{16} + \frac{(2p-1)}{p}\left( \frac{2\cdot 5}{16} +\frac{3\cdot 7}{16}\right)\right] \\
&- \left(p^2(1-p)\frac{2p-1}{p} + \frac{p(1-p)^2}{8}\frac{2p-1}{p}\right) \frac{3}{4}\\
=& \frac{(2p-1)(3p+1)^3}{6(5p-1)} - \frac{(2p-1)(63p^2 + 11p + 3)}{32} - \frac{(2p-1)(1-p)(14p+3)}{32} - \frac{3(2p-1)(1-p)(7p+1)}{32}\\
=& (2p-1)\left[ \frac{(3p+1)^3}{6(5p-1)} - \frac{28p^2 +40p + 9}{32}\right]. 
\end{split}
\end{equation*}

\section{Derivation of the approximated auto-correlation functions of time differences $k = 2$ and $3$}
\label{appendix-appro-corre23}
\begin{equation*}
\begin{split}
\langle x_2 x_0 \rangle \simeq & p^2\int_0^1 \rho_p (x_0)(4x_0 \mod 1)x_0 dx_0 + 2p(1-p)\int_0^1 \rho_p(x_0) x_0^2 dx_0 + (1-p)^2 \int_0^1 \rho_p(x_0)\frac{1}{4}x_0^2 dx_0\\
=& p^2 \left[ \int_0^{\frac{1}{4}}\rho_p(x_0)4x_0^2 dx_0 + \int_{\frac{1}{4}}^{\frac{1}{2}} \rho_p(x_0)(4x_0 - 1)x_0dx_0 + \int_{\frac{1}{2}}^{\frac{3}{4}} \rho_p(x_0) (4x_0 - 2)x_0dx_0 \right.\\
& \left. + \int_{\frac{3}{4}}^1 \rho_p(x_0)(4x_0 - 3)x_0dx_0\right] + 2p(1-p)\langle x^2\rangle + \frac{(1-p)^2}{4}\langle x^2\rangle \\
=& p^2 \left[ \int_0^1 \rho_p(x_0)4x_0^2 dx_0 - \int_{\frac{1}{4}}^{\frac{1}{2}} \rho_p (x_0)x_0 dx_0 - 2\int_{\frac{1}{2}}^{\frac{3}{4}} \rho_p(x_0)x_0 dx_0 - 3\int_{\frac{3}{4}}^1 \rho_p(x_0)x_0 dx_0\right] + \left[ 2p(1-p) + \frac{(1-p)^2}{4}\right] \langle x^2\rangle\\
=& \left[ 4p^2 + 2p(1-p) + \frac{(1-p)^2}{4}\right] \langle x^2\rangle - p^2\left[ a_1 \int_{\frac{1}{4}}^{\frac{1}{2}} x dx + 2a_0 \int_{\frac{1}{2}}^{\frac{3}{4}} x dx + 3a_0 \int_{\frac{3}{4}}^1 x dx\right]\\
=& \frac{(3p+1)^2}{4}\langle x^2 \rangle - p^2 \left[ \frac{(2p-1)(2-p)}{p^2}\cdot\frac{3}{32} + 2\cdot \frac{2p-1}{p}\cdot \frac{5}{32} + 3\cdot \frac{2p-1}{p}\cdot \frac{7}{32}\right]\\
=& \frac{(3p+1)^2}{4}\cdot \frac{4(2p-1)}{3(5p-1)} - \left[ \frac{3(2p-1)(2-p)}{32} + \frac{5p(2p-1)}{16} + \frac{21p(2p-1)}{32}\right]\\
=& (2p-1)\left[ \frac{(3p+1)^2}{3(5p-1)} - \frac{14p+3}{16}\right]. 
\end{split}
\end{equation*}
The third iteration of the approximated system is given by 
\begin{equation*}
x_3 = \begin{cases} 
T_1(T_1(T_1 (x_0))) = 2^3 x_0 \mod 1, &\text{prob. $p^3$}\\
T_i(T_j(T_k (x_0))) = 2x_0 \mod 1, \quad \text{only one of $i, j, k$ is $1$}, &\text{prob. $3p^2(1-p)$}\\
T_i(T_j(T_k (x_0))) = \frac{1}{2}x_0, \quad \text{only one of $i, j, k$ is $2$}, & \text{prob. $3p(1-p)^2$}\\
T_2(T_2(T_2 (x_0))) = \frac{1}{2^3}x_0, &\text{prob. $(1-p)^3$}, 
\end{cases}
\end{equation*}
and 
\begin{equation*}
\begin{split}
\langle x_3 x_0 \rangle \simeq & p^3 \int_0^1 \rho_p(x_0) (8x_0 \mod 1)x_0 dx_0 + 3p^2(1-p)\int_0^1 \rho_p(x_0) (2x_0 \mod 1)x_0 dx_0 + 3p(1-p)^2\int_0^1 \rho_p(x_0) \frac{1}{2}x_0^2 dx_0\\
&+ (1-p)^3\int_0^1 \rho_p(x_0) \frac{1}{8}x_0^3 dx_0\\
=& p^3 \left[ \int_0^{\frac{1}{8}}\rho_p(x_0) 8x_0^2 dx_0 + \int_{\frac{1}{8}}^{\frac{2}{8}}\rho_p(x_0) (8x_0 - 1)x_0 dx_0 + \int_{\frac{2}{8}}^{\frac{3}{8}}\rho_p(x_0) (8x_0 - 2)x_0 dx_0 + \int_{\frac{3}{8}}^{\frac{4}{8}}\rho_p(x_0) (8x_0 - 3)x_0 dx_0\right. \\
& \left. \int_{\frac{4}{8}}^{\frac{5}{8}}\rho_p(x_0) (8x_0 - 4)x_0 dx_0 + \int_{\frac{5}{8}}^{\frac{6}{8}}\rho_p(x_0) (8x_0 - 5)x_0 dx_0 + \int_{\frac{6}{8}}^{\frac{7}{8}}\rho_p(x_0) (8x_0 - 6)x_0 dx_0 + \int_{\frac{7}{8}}^1 \rho_p(x_0) (8x_0 - 7)x_0 dx_0 \right]\\
& + 3p^2(1-p)\left[ \int_0^{\frac{1}{2}}2x_0^2 dx_0 + \int_{\frac{1}{2}}^1(2x_0 - 1)x_0dx_0\right] + \frac{3p(1-p)^2}{2}\langle x^2 \rangle + \frac{(1-p)^3}{8}\langle x^2 \rangle \\
=& p^3 \left[ 8\int_0^1 \rho_p(x_0)x_0^2 dx_0 - \int_{\frac{1}{8}}^{\frac{2}{8}} \rho_p(x_0)x_0dx_0 - 2\int_{\frac{2}{8}}^{\frac{3}{8}} \rho_p(x_0)x_0dx_0 - 3\int_{\frac{3}{8}}^{\frac{4}{8}} \rho_p(x_0)x_0dx_0 - 4\int_{\frac{4}{8}}^{\frac{5}{8}} \rho_p(x_0)x_0dx_0 \right. \\
&\left. - 5\int_{\frac{5}{8}}^{\frac{6}{8}} \rho_p(x_0)x_0dx_0 - 6\int_{\frac{6}{8}}^{\frac{7}{8}} \rho_p(x_0)x_0dx_0 - 7\int_{\frac{7}{8}}^1 \rho_p(x_0)x_0dx_0 \right] + 3p^2(1-p)\left[ 2\int_0^1\rho_p(x_0)x_0^2dx_0 - \int_{\frac{1}{2}}^1\rho_p(x_0) x_0dx_0\right] \\
&+ \left[ \frac{3p(1-p)^2}{2} + \frac{(1-p)^3}{8}\right]\langle x^2 \rangle \\
=& \left[ 8p^3 + 3p^2(1-p) + \frac{3p(1-p)^2}{2} + \frac{(1-p)^3}{8}\right] \langle x^2 \rangle - p^3 \left[ a_2\int_{\frac{1}{8}}^{\frac{2}{8}}xdx + 2a_1\int_{\frac{2}{8}}^{\frac{3}{8}}xdx + 3a_1\int_{\frac{3}{8}}^{\frac{4}{8}}xdx + 4a_0\int_{\frac{4}{8}}^{\frac{5}{8}}xdx \right. \\
&\left. + 5a_0\int_{\frac{5}{8}}^{\frac{6}{8}}xdx + 6a_0\int_{\frac{6}{8}}^{\frac{7}{8}}xdx + 7a_0\int_{\frac{7}{8}}^1xdx\right] - 3p^2(1-p)a_0\int_{\frac{1}{2}}^1xdx\\
=& \left(2p + \frac{(1-p)}{2}\right)^3\langle x^2 \rangle - p^3 \left[ \frac{(2p-1)(3p^2 - 6p + 4)}{p^3}\cdot \frac{3}{128} + 2\cdot \frac{(2p-1)(2-p)}{p^2}\cdot \frac{5}{128} + 3\cdot \frac{(2p-1)(2-p)}{p^2}\cdot \frac{7}{128} \right. \\
&\left. + 4\cdot \frac{2p-1}{p}\cdot \frac{9}{128} + 5\cdot \frac{2p-1}{p}\cdot \frac{11}{128} + 6\cdot \frac{2p-1}{p}\cdot \frac{13}{128} + 7\cdot \frac{2p-1}{p}\cdot \frac{15}{128}\right] - 3p^2(1-p)\cdot \frac{2p-1}{p} \cdot \frac{3}{8}\\
=& \frac{(3p+1)^3}{8}\frac{4(2p-1)}{3(5p-1)} - \left[ \frac{3(2p-1)(3p^2 - 6p + 4)}{128} + \frac{10p(2p-1)(2-p)}{128} + \frac{21p(2p-1)(2-p)}{128} + \frac{36p^2(2p-1)}{128} \right. \\
&\left. + \frac{55p^2(2p-1)}{128} + \frac{78p^2(2p-1)}{128} + \frac{105p^2(2p-1)}{128} \right] - \frac{9p(1-p)(2p-1)}{8}\\
=& (2p-1) \left[ \frac{(3p+1)^3}{6(5p-1)} - \frac{63p^2 + 11p + 3}{32} - \frac{9p(1-p)}{8}\right] \\
=& (2p-1) \left[ \frac{(3p+1)^3}{6(5p-1)} - \frac{27p^2 + 47p + 3}{32} \right]. 
\end{split}
\end{equation*}

\bibliography{rdsbib}

\end{document}